\documentclass[12pt,a4paper,oneside]{amsart}

\usepackage{amsfonts, amsmath, amssymb, amsthm, hyperref}
\usepackage{anysize}

\usepackage[T2A]{fontenc}
\usepackage[cp1251]{inputenc}
\usepackage[english]{babel}
\newtheorem{theorem}{Theorem}[section]
\newtheorem*{theorem*}{Theorem}

\theoremstyle{definition}

\theoremstyle{remark}

\numberwithin{equation}{section}

\renewcommand{\epsilon}{\varepsilon}
\renewcommand{\phi}{\varphi}
\renewcommand{\kappa}{\varkappa}

\begin{document}

\title{Circuit Satisfiability Problem for circuits of small complexity}
\author{Marsel~Matdinov}

\email{}
\address{}


\begin{abstract}
The following problem is considered. A Turing machine $M$ that accepts a string of fixed length $t$ as input,  runs for a time not exceeding a fixed value $n$ and is guaranteed to produce a binary output is given. It's required to find a string $X$ such that $M(X) = 1$ effectively in terms of $t$, $n$, the size of the alphabet of $M$ and the number of states of $M$. The problem is close to the well-known Circuit Satisfiability Problem. The difference from Circuit Satisfiability Problem is that when reduced to Circuit Satisfiability Problem, we get circuits with a rich internal structure (in particular, these are circuits of small Kolmogorov complexity). The proof system, operating with potential proofs of the fact that, for a given machine $M$, the string $X$ does not exist, is provided, its completeness is proved and the algorithm guaranteed to find a proof of the absence of the string $X$ in the case of its actual absence is presented (in the worst case, the algorithm is exponential, but in a wide class of interesting cases it works in polynomial time). We present an algorithm searching for the string $X$, for which its efficiency was neither tested, nor proven, and it may require serious improvement in the future, so it can be regarded as an idea. We also discuss first steps towards solving a more complex problem similar to this one: a Turing machine $M$ that accepts two strings $X$ and $Y$ of fixed length and running for a time that does not exceed a fixed value is given; it is required to build an algorithm $N$ that builds a string $Y = N(X)$ for any string $X$, such that $M(X, Y) = 1$ (details in the introduction).
\end{abstract}

\maketitle

\sloppy


\tableofcontents

\newpage

\section{Introduction.}

We will talk about a problem that is close to the question of equality of classes $P$ and $NP$. I assumed that many problems of olympiad programming, discrete mathematics, and theoretical computer science can be formulated as special cases of the following problem. Turing machine $M$ that accepts two strings is given. (it will be convenient for us to solve the problem if the lengths of these two strings are fixed and the machine is running for a time that does not exceed a fixed value). It's required to build a Turing machine $N$ that builds either a string $N(x)$, given a string $x$, such that $M(x, N(x)) = 1$, or outputs a special character if there is no such string $y$ that $M(x, y) = 1$. We will refer to this problem as the "big string writing problem".

As a "small string writing problem"\ we will call the following problem. A Turing machine is given that accepts a fixed-length string as input and runs for a time not exceeding a fixed value. It's required to write a string on which this Turing machine returns 1 or make sure that there is no such string.

A small string writing problem can be reduced to a big one, a big problem can be almost reduced to a small one. I do not want to talk now about these simple reductions and explain what "almost"\ means, we will not need it yet.

There is a large, pretty poorly defined, class of problems similar to these two which are often found in mentioned areas. For example, the following problem: Given a Turing machine (hereinafter - TM) $M$ that accepts three strings. It is required to build a Turing machine $N$ that builds either a string $N(x)$ such that $M(x, N(x), y) = 1$ for any string $y$, or outputs a special character if there is no such string $t$ that $M(x, t, y) = 1$ for any string $y$ (this time the length of the string $y$ is not limited).

To solve one of these problems, it is reasonable to solve all these problems at once, because solving one we can meet the other as a subtask.

Of course, all these problems are algorithmically unsolvable in general case. But, as I said, in practice, very often people successfully solve problems of this type.

Firstly it should be said that, first of all, we are looking for an efficient algorithm. For example, in the case of small string writing problem, we will not be satisfied by algorithm that, roughly speaking, iterates through all the strings and tries to substitute them in the given TM. Secondly, in the case of a big string writing problem, we want to see as an answer, not any algorithm, but, again, an effective one, otherwise it would give the opportunity to construct an algorithm that iterates over all possible candidates for the role of $N(x)$ and somehow compares them. In other words, in the big string writing problem, our goal is to efficiently search for an efficient algorithm.

Let's talk about small string writing problem. We assume that the TM's head is initially located in the first cell of the input string. Let's also agree that at the end of work of TM, the head of our TM comes to the same first cell, where it was initially, and one of the TM states at the end of the work encodes the output; and let's agree that after the end of TM's work, neither the symbols of the tape, nor the position of the head, nor the state of the TM is changed anymore. If the TM given to us does not meet these conditions, it is easy to change it so that it meets them. The time of its work will not increase much. Calculation of the output by described, working during time, non exceeding fixed amount of time, TM, in a standard well known way can be converted to calculating the output from the same string using a boolean circuit. This calculation is arranged as follows. Note that since the machine runs for a time not exceeding $n$, its head moves to a distance no more than $n$ from its initial position during operation. Consider the cell rectangle $(2n + 1) \cdot n$. Let's identify its lower side with the segment of tape on which the TM is working, in its initial state. The left and right ends of this segment are at a distance of $n$ from the initial position of the head. Consider running of our TM on some input string. Let's mark the cell of the rectangle with the coordinates $(x, y)$ with the state of the cell of the tape with the number $x$ at the moment of time $y$. The state of the cell is made up of the character, written in this cell at the moment, a bit that means the presence of the TM head in this cell at the moment, as well as the state of the TM at a given moment in time. All three objects are encoded with zeros and ones. Note that the label of a cell with coordinates $(x, y)$ is determined by the labels of cells with coordinates $(x-1, y-1)$, $(x, y-1)$, $(x + 1, y- 1)$ (we work in a computation model where there is one head, at any time we look at the symbol in the tape where the head is situated and at the state of TM; depending on them, we choose the next state, change the symbol at the current cell of the tape and also choose whether the head will remain in place, move to the left or move to the right). Thus, we can construct a boolean circuit, each gate of which corresponds to the bit of a label of certain rectangle cell, and the value of this gate is calculated naturally from all the gates of a labels of three rectangle cells located on a row one lower, directly below this cell, slightly to the left or slightly to the right. The line that we feed to the input of our TM determines the input of the circuit, and the value of one of the gates corresponding to the cell on the upper side of the rectangle, which corresponds to the tape cell, on which the first character of the input line was located at the beginning of the TM operation, is the output that the circuit gives.

In most sections, we will focus on the small string writing problem, which, as we have just seen, is equivalent to the problem of selecting the input values of a special Boolean circuit, such that this circuit outputs 1. We will often refer to this variant of formulation with the circuit as a small string writing problem.

If we do not impose any restrictions on the Boolean circuit for which we want to find an input, then we get the Circuit Satisfiability Problem - the well-known $ NP $ -complete problem. Sinse it is believed that $P \neq NP$, we will hardly be able to come up with a polynomial algorithm for such a problem. Our case differs in that the circuit has a certain structure. In our case the circuit has a small Kolmogorov complexity.

In the case of a big string writing problem, we can similarly "restrict the computation to a rectangle"\ , two strings of fixed length (separated by a separator) will be fed to this rectangular circuit, and our task is to build an algorithm that, based on the first of these strings (to the left of the separator) constructs a suitable second row (to the right) such that the circuit outputs 1 on this pair of strings.

Unfortunately, now I can say little about class of problems the development of my algorithm is aimed at. The algorithm is stochastic and it will hardly be possible to outline this class explicitly and strictly. However, if our algorithm does not solve some unnatural, fanciful problem like "given a graph, you need to place different prime numbers on its edges so that the sum of the numbers on each cycle of length no more than 7 is represented as $n^3 + n + 2$ for the integer $n$"\ , then, I think, it can be forgiven for this. (The given example of the problem is an instance of the big string writing problem - the given properties of the arrangement of numbers on the edges are checked on a polynomial Turing machine.) It would be nice to solve a problem in approximately those cases in which a human copes with it.

We can also consider a following generalization of the big string writing problem. It is required, given TM $M$ to construct an algorithm $A$, which, given the strings $X$, $Y$, constructs a string $Z$ such that $M (X, Y, Z) = 1$. The difference is that we cannot see the $Y$ string, it is, so to speak, an invisible part of the world. But we are allowed to experiment with this invisible string. That is, we are given another TM $N$, which takes this invisible string $Y$ and any string that the algorithm $A$ wants, giving us an output. The work of the algorithm $A$, therefore, consists, among other things, of generating requests to this very machine $N$ and processing the outputs it gives.

A similar task to the one under consideration is given by Marcus Hutter in the book ~\cite {GP2007} in the section "The Cybernetic Agent Model"\ . In short, in it the agent represented by TM $p$ interacts with the environment represented by TM $q$. The interaction consists of cycles, in the $i$-th of which, based on the interaction history $y_1 x_1 ... y_{i-1} x_{i-1}$, the agent issues the string $y_i$, after which the environment, based on the same interaction history, supplemented by the string $y_i$, issues the string $x_i$. The string $x_i$ always consists of two parts: the main part $o_i$ and the reward $r_i$. Our goal is to build an agent $p$ that maximizes the sum of $r_i$. TM $q$ can be probabilistic. The book examines the problem in the aspect of reinforcement learning. However, the big string writing problem can be almost thought of as a special case of this one. Namely, if it is the environment that makes the first (probabilistic) move, after which the agent makes the move, after which the environment makes its second, final deterministic move containing only a reward, that is equal to one, if the agent's move in combination with the first move of the environment satisfies it and is equal to zero if not satisfied. The environment, in this case, must be known for us.

\section {Proof system}

\subsection {Ordinary}

We will talk about the circuits that implement the computation on the Turing machine, described in the previous section. It would be nice to certify the fact that such a circuit issues 0 at all possible inputs. In other words, to build a construction, the presence of which indicates that the circuit always outputs 0. This proof system can be naturally generalized to circuits of a somewhat more general form, namely, circuits on which there is a naturally defined notion of proximity of gates and for each gate, you can select a small neighborhood of this gate, in which the number of gates is constant.

We assign each gate (including input and output) an arbitrary value - 0 or 1, but the gate, which is considered to be the output produced by the circuit, we assign 1. We want to prove that any such set of values is inconsistent, that is, there is a gate that is not fulfilled by this a set of values. (The expression "gate is not fulfilled by a set of values" means that the value of this gate does not correspond to the value assigned to this gate by the values of the gates on which it depends).

The proof is an object that we call "cage". A cage is a collection of small sets (of constant size) of closely spaced gates (from some constant neighborhood), to each of which a function is tied, defined on a constant number of boolean variables, which assigns a rational number to a set of values in these gates. At that, two statements must be satisfied:

1) The sum of the values of all these functions does not depend on the values that we bind to the gates and it is greater than zero (let me remind you that in a gate that is considered an output, there is always 1, it cannot be changed). This property is quickly checked; it is enough to check it locally: only those few functions depend on the value of given gate, which depend on several gates from the vicinity of this gate, including this gate itself (localized in the vicinity of this gate). One of the required properties of this collection of functions, which I, however, did not highlight as a separate item, is that there is only constant number of functions that depend on one or several gates from a constant neighborhood of any particular gate (neighborhood includes this gate itself) (however, if there were a lot of such functions for some gate, then those of them that depend on the same gates, we can simply sum up and replace them all by the sum). And to check, we can simply iterate over the values of all the gates from some sufficiently large but constant neighborhood and check that the sum of functions depending only on the gates of the neighborhood does not change when we change the value of the considered gate from zero to one and back.

If we check this for each gate, then the invariance will be proved, since any set of gate values can be translated into any other by changing the values one by one. To make sure after that that the sum of the values of the functions is greater than zero, it is enough to check this for an arbitrary set of gate values.

2) From the presence of a function of a cage for a given set of values of gates, which is positive on this set of values, it follows that in some constant neighborhood of the set of gates on which this function depends, a gate that is not fulfilled by this set of values exists. This property is checked by iterating, for each function, through the all sets of values associated with the gates of the corresponding neighborhood and finding, for each set of values on which the function is positive, a gate in this neighborhood that is not fulfilled. Perhaps this neighborhood should be indicated in the cage itself (in the presented structure itself).

Obviously, from the fulfillment of these two conditions, it follows that the circuit never outputs one: if it outputs one, we would consider a set of values tied to the gates, at which all gates are fulfilled and with it, the sum of the function values would not be positive, by property (2) , and would be positive by property (1), a contradiction.

\subsection{With suspension and additional construction}

Now let's talk about the "generalized cage". We will consider two improvements over the ordinary cage. The first is that we are introducing additional gates and additional restrictions on gates. (There were restrictions that the value of each of the old gates is equal to a function of the values of the old gates located directly below it in the circuit, now there are new gates and new restrictions. Each of the new restrictions is a prohibition for a constant number of gates to take a certain set of values.) At the same time, it must be fulfilled that if in the old circuit (without additional gates) there is a fulfilling set of gate values, then it can be supplemented with the values of the gates added to the circuit so that all new constraints are also satisfied. This can be achieved most simply by requiring that there is an order on the gates, in which the old gates occupy positions from 1 to $n$, and the added gates occupy positions from $n + 1$ to $n + m$, where $n$ is the number of old gates, and $m$ is the number of gates added; and $m$ restrictions are added, the $i$ -th of which says that the value of the $n + i$-th gate is equal to the function of several previous gates, and such that it implies that the $n + i$-th gate does not conflict with the previous gates (is not contained in forbidden configuration with previous gates that is other than $m$ indicated restrictions). That is, we sort of "build up" the circuit outward. So if we have the proper values in the old gates, we can simply compute the values in the new gates one by one, according to the new constraints.

In the general case, such a construction will not be a circuit, but rather a set of clauses, so we will call it a generalized circuit. Also, a generalized circuit is required to have a system of constant neighborhoods in it. It is needed so that the invariance of the sum of the values of all functions can be checked locally (we will consider the system of functions on the gates of a generalized circuit in absolutely the same way). Formally, this requirement can be formulated by presenting the definition of a $d$-constant generalized circuit. Let us fix some constant number $d$. The $d$-constant generalized circuit must satisfy the following conditions:
1. It is a metric space, that is, for each pair of gates a real non-negative distance between them is defined, which satisfies the properties of a metric space (identity of indiscernibles, symmetry, triangle inequality).
2. Each $d$-neighborhood of a gate contains only a constant number of gates.
3. If two gates are included in the same clause, then the distance between them is not more than 1.

We will work with $d$-constant circuits, again, in order to be able to check the invariance of the sum of the values of all functions locally.

A good example of a $d$-local circuit is a rectangular parallelepiped extended from above to the rectangle of the original circuit, as on a base, each gate of which depends on those gates of its neighborhood that are located slightly below it, or slightly to the left of it but not above it (we assume that the original circuit is calculated from left to right). For such a class of generalized circuits, the choice of $d$ is not essential.

The second improvement is as follows. We will not have one cage, but a tree of cages. Each of them is, in the same way, a set of functions on the values of the gates, each of which depends on a constant number of gates and these gates must be located next to each other. The first difference (though not essential) from functions from the old version of cage is that each of these functions is equal to one on exactly one of the sets of values of gates of its domain. On any other set of values of this gates, it is zero. That is, a function is simply a light bulb that lights up if and only if the gates on which it depends take a strictly defined set of values (hereinafter we will sometimes call them that way - light bulbs). The second difference is that a rational weight is assigned to each such function. Moreover, each of the cages of the tree (and, thus, all of its functions) is not defined on the entire set of values of all gates, but only on those where certain fixed values are assigned to some gates. In other words, for each cage, some gates are constant.

The first condition for such a cage is that the sum of the weights of the lit light bulbs is invariant on the set of sets of gate values, in which the correct values are assigned to the constant gates, and is equal to some number greater than zero.

Next, how will our tree be arranged in general? It will be suspended from the top vertice. The top cage will have no constant gates. The cages that are suspended from this cage will correspond to the light bulbs of this cage, which have a positive weight. In each of these cages, several constant gates appear - exactly those on which the corresponding light bulb depends, and for these gates, exactly that values will be assigned which are needed for the corresponding light bulb to light up. The cages of the third level suspended from the cage of the second level correspond to the light bulbs of this cage of the second level of positive weight and their constant gates are constant gates of the corresponding cage of the second level (with the same values), to which the gates on which the corresponding light bulb depends, set exactly at those values that are needed to this light to light up, are added. And so on. The vertices of the tree correspond to the cages, the edges correspond to the light bulbs of these cages with a positive weight. At the same time, each positive light bulb of each cage of the tree can correspond to either a cage of the level one lower (as it was just described), or a specific restriction, which is violated by the gate values that make this light lighting up (this time the gates included in this restriction, all must be inside the set of gates of domain of this light bulb, and not just in the neighborhood). Let's call the light bulb of the second kind (breaking the constraint) terminal.

So, how do we find the violated gate by the gate values and the tree of cages? Let's look at the top cage. Some of its light bulbs light up. The sum of their weights is greater than zero. Let us choose from the lit light bulbs the one that has positive weight. It corresponds to a cage one level below (or a terminal light). Let's choose a positive lit light bulb in it. And so on. Sometime we will come to a lit terminal light bulb and a violated gate contained in its domain.

The invariance of sum of weights of lit light bulbs of each cage of the tree is checked, again, locally, for each non-constant gate. However, since we are working with a $d$-constant generalized circuits, it is worth clarifying that the diameter of the domain of light bulb should not exceed $d / 3$ (then we will definitely be able to check the invariance locally, for each cage of the tree).

It should be said that I have analyzed several (nearly eight) examples of circuits of different (mostly not big) complexity that cannot output 1, and in all of them, except the last two, I managed to come up with a cage. Just a cage, not a generalized cage. This gives me a reason to assume that the proof system "cage" is quite strong.

For the last two examples, I also managed to come up with a proofs, but they required the addition of new gates to the circuit and a tree of cages of nontrivial depth.

It was nice that among the examples I analyzed there was a circuit that encodes the Dirichlet principle: the inputs of this circuit form a rectangle $n \cdot (n + 1)$, this rectangle is adjoined from above by $n + 1$ circuits that check that each column contains at least one 1, as well as $n$ circuits checking that each line contains at most one 1. Higher, conjunction is made with the results of these $2n + 1$ circuits. It is clear that no matter how we put zeros and ones in the inputs, the output of the circuit (the result of conjunction) will be zero. I managed to build a cage for this circuit - to arrange the bulbs on this circuit so that the sum of the weights of ones lit would always be one and each lit light bulb of positive weight concealed a violated restriction. It was pleasant for me, because, in particular, it allows us to prove various estimates and inequalities - in my opinion, a rather difficult thing to automate.

I want to believe that for problems that arise in practice, it is often possible to come up with a generalized circuit (of course, if the corresponding TM never outputs one), which, moreover, has a small complexity and a small number of additional gates.

\section {Links to topology}

The described proof system arose from topological considerations and has a topological nature. I will not describe this connection with the topology in details, since we will not need it, I will only say in short what it is about. This can be seen on the example of a simple circuit that comes from the natural decomposition of a large equilateral triangle into small equilateral triangles of the same size (the circuit does not come from the work of a Turing machine). We place a gate of circuit in each node of this partition. We will arrange the triangle so that one of its vertices is at the top, and the side, opposite to it, is horizontal (and the input gates of the circuit are located at the nodes of the partition on this side). The value at each node will somehow be calculated from the two values at the nodes directly below this node (and forming a regular triangle with it). We embed this plane, divided into triangles, into three-dimensional space and put a point at each node of the partition. Let us draw from each set point the normal to the plane of length 1 in the same direction and put a point at the end of each such normal. The point at the start of the normal corresponds to the value 0 in the corresponding gate, the point at the end of the normal corresponds to the value 1 in this gate.

Let's call the nodes on the border of the triangle, as well as the points obtained from these nodes by adding of the indicated normal, boundary points. Let us embed our three-dimensional space into a seven-dimensional space. We will slightly move each marked point so that in seven-dimensional space they form points of general position. Consider two adjacent nodes of the boundary of the original triangle, as well as two points that differ from them by the indicated normal. Consider the 4 points that came out of them after the shift. In seven-dimensional space, they form a tetrahedron. Let's call it boundary tetrahedron. Let us call the union of all boundary tetrahedra a boundary of the circuit.

Consider an arbitrary cage that assosiates each small triangle with binary values fixed at its vertices with a rational number, the weight of a light bulb corresponding to a given small triangle with values fixed at its vertices - exactly these values are needed for a bulb to light up (in this section, we will talk about just such cages - cages, each of whose bulbs depends exactly on the values at the vertices of the small triangle). Using the property of invariance of the sum of the weights of the bulbs lit up in the cage, I managed to construct an oriented five-dimensional manifold with a zero boundary, the intersection index of which with each triangle, the vertices of which are a triple of shifted nodes projected approximately to the vertices of a small equilateral triangle (to the neighborhood of one vertex of a small equilateral triangle, let me remind you, two shifted nodes are projected, differing one from another by the normal discussed above) is exactly equal to the weight of the light bulb corresponding to this triple. Moreover, this manifold does not intersect the boundary of the circuit. In addition, consider an arbitrary choice of zeros and ones at the circuit nodes and consider a two-dimensional manifold equal to the union of the triangles corresponding to the lit light bulbs (this is a "two-dimensional film" whose boundary lies on the circuit's boundary). So, the intersection index of our five-dimensional manifold with such a film is greater than zero (and from the fact that our five-dimensional manifold with zero boundary does not intersect the boundary of the circuit, and any film, obviously, can be continuously deformed into any other such film (corresponding to another set of zeros and units) so that the boundary of the deformed film always remains at the boundary of the circuit, it follows that the indicated intersection index is the same for all such films (does not depend on the choice of zeros and ones)).

The converse is also true. Given a five-dimensional manifold without an edge that does not intersect the boundary of the circuit, of a suitable homological class (that is, intersecting the above-described film of triangles corresponding to an arbitrary choice of zeros and ones, in a positive way), one can construct a cage, provided that this five-dimensional manifold intersects each triangle corresponding to light bulb "in the right way". The phrase "in the right way"\ means the following. We have a condition for the cage that some of its bulbs can have an arbitrary weight, and the rest - not a positive weight. So, our variety is allowed to intersect triangles corresponding to bulbs of the first type in an arbitrary way, and triangles corresponding to bulbs of the second kind - not in a positive way. Let's agree that for our manifold it is forbidden to cross the sides of the triangles, this does not limit us too much. A cage is constructed from a manifold in an obvious way: the index of intersection of the manifold with the corresponding triangle is taken as the weight of the bulb.

Thus, in the case of this circuit and the considered family of cages (cages operating with "triangular" \ bulbs), the search for a cage that is a correct proof is reduced to a search for a manifold in a certain space of the desired homological class (positively intersecting a two-dimensional film "stretched"\ onto the boundary of a circuit) passing through this space "in the correct way"\ (some of the triangles can be crossed only "in one direction"). 

When constructing a manifold from a cage, we used one structure, the construction of which, in particular, required that the dimension of the enclosing space be at least 7. I almost succeeded in completing of a similar construction (manifolds from a cage and vice versa) for the case of an arbitrary circuit that come from a Turing machine. Now I see no reason to tell in detail what "almost"\ means, I will only say that the existence of a manifold in the general case, given by my construction, depends on whether a certain manifold always contracts homologically within a certain space. I have not verified this fact yet, but it looks like it is true. If it is true, my construction works, and if not, then it will be necessary to complicate the construction a little, and then the ambient space (in which we are looking for a manifold) will not be Euclidean. One way or another, if someone needs it, I'm ready to refine the construction for the case of arbitrary circuits that came from TM.

\section{Cocyclic polynomials}

\subsection{Definition}

We will talk about the generalization of the proof system "cage". A cage, ordinary or "multistorey"\ (when there is a tree of cages) illustrates this proof system very well. To the case when we add additional "external"\ gates to the circuit, just as it was when we described the system "generalized cage"\, this system can also be generalized in an obvious way.

So, let's consider the circuit that came from the work of TM. For the gates of this circuit we can naturally determine the distance. To do this, you can, for example, naturally embed the circuit in a 3-dimensional space. To do this, each gate of the circuit, representing a certain bit of the state of a certain cell of tape of number $x$ at the time $y$, is associated with a point in three-dimensional space near the point ($x$, $y$, 0). And the distance between the corresponding points we declare the distance between the gates.

Consider the constant $k$. Our proof system will operate with the same light-bulb-functions that were in the "cage" system. Each light bulb lights if the gates it depends on take a specific set of values. The diameter of a light bulb is the maximum distance between the gates on which it depends. We will be interested in light bulbs with a diameter of no more than $k$. We will call such bulbs small.

Consider an integer $d$ (usually, $d$ will be a constant). Let's define a cocyclic polynomial. Cocyclic polynomial of degree $d$ is a polynomial of degree $d$, the variables of which are in one-to-one correspondence with the small light bulbs of our circuit and which satisfies the invariance condition. The invariance condition is that if we choose the values of gates of the circuit in an arbitrary way and substitute in each variable the value of the corresponding light bulb (0 if the bulb is not lit and 1 if the bulb is lit), then the value issued by the polynomial will be independent of the choise of the gate values.

A cocyclic polynomial can be considered a proof of the insatisfiability of a circuit if it satisfies the security condition. A cocyclic polynomial is considered satisfying the security condition if 2 conditions are met. First, the value it produces for any choice of gate values (the same for all choices of values) is greater than zero. And secondly, for each monomial with a positive coefficient, at least one of the bulbs corresponding to its variables "implies a contradiction." The phrase "implies a contradiction"\ means that among the gates on which the value of this bulb depends, there is some gate $a$ and all the gates on which $a$ depends in the circuit, and at the same time, if these gates take the values that are needed for the bulb to light up, then they will form a contradiction: the value in the gate $a$ will not correspond to the value prescribed to $a$ by the circuit according to the gate values on which $a$ depends in the circuit. (In theory, nothing would prevent us from also allowing the coefficients of monomials to be positive, for which it's union of domains of bulbs of the corresponding monomial with corresponding values ascribed to each of its gates (so that each bulb of monomial lights up) that implies a contradiction - a slightly weaker condition.)

Indeed, if a circuit with a cocyclic polynomial satisfying the security condition (hereinafter referred to as a protected cocyclic polynomial) were satisfiable, we would take a set of gate values satisfying it, choose a monomial of a cocyclic polynomial with a positive coefficient that does not equal to zero (it exists, since if it did not exist, the value of the polynomial for a given set of gate values would not be positive), we would choose a light bulb corresponding to one of its variables, implying a contradiction, and extract from it the gate violated by this set of values.

Note that for $d = 1$ we get exactly the proof system "cage".

By a lamp polynomial of degree $d$ we mean a polynomial of degree $d$, whose variables are in one-to-one correspondence with the small bulbs of our circuit (which, however, does not have to satisfy the invariance condition). The lamp polynomial can be viewed as a function of choise of the gate values of the circuit, calculated in the same way as described in the paragraph just above (just like in the case of a cocyclic polynomial).
 
\subsection{Local check for cocyclicity.}

The good news are that the cocyclicity of a polynomial can be checked locally. Consider a polynomial $f$ of degree at most $d$. Let's check if it is cocyclic. Let's choose the gate $u_1$. Let's check that regardless of the values of all other gates, the value of $f$ does not change when the value of $u_1$ changes from zero to one. Consider a neighborhood of $u_1$ consisting of all gates at a distance of at most $k$ from it. Since $k$ is a constant, we can iterate over all possible values of all gates in this neighborhood except for $u_1$. Let's do it. And for each set of values of these gates, by temporarily fixing it, we will check that for any possible set of values of all other gates, the value of $f$ does not change when $u_1$ is changed from zero to one.

The key idea is that for fixed values of the neighborhood gates other than $u_1$, the change in the value of $f$ when the value of $u_1$ changes from zero to one is a lamp polynomial, with light bulbs depending on all other gates, of degree which is at least one less. Why is this so? Consider an arbitrary monomial $f$. It is a product of several bulbs (to be more precise, the variables corresponding to the bulbs), with a certain coefficient. There are two possibilities.

The first possibility is that none of these bulbs "cover"\ $u_1$ (more precisely, none of these bulbs depend on $u_1$). This means that the value of this monomial does not depend on the value of $u_1$ and, thus, contribution of this monomial to the considered change in the value of the polynomial, when the value of $u_1$ changes from zero to one, is zero.

The second option is that $u_1$ is covered by at least one monomials bulb. Note that in this case the monomial value can be nonzero only for $u_1 = 0$, or only for $u_1 = 1$ (because the bulb containing $u_1$ can take a nonzero value, as a maximum, in one of these two options). And this is only if the values at which the monomial bulbs light up are consistent with the values we fixed in the "punctured" neighborhood of $u_1$: if at least one of the gate values needed for one of these bulbs to light up does not coincide with the fixed value of the corresponding gate of the punctured neighborhood $u_1$, then the value of the monomial in both cases will necessarily be equal to zero.

Thus, we are interested in precisely this subcase of the second case, when the values at which the monomial bulbs light up are consistent with the values we fixed in the punctured neighborhood of $u_1$.

Let's call the truncated monomial bulb the monomial bulb, from which all dependences on the gates included in the considered neighborhood of $u_1$, including $u_1$, have been removed. Let me explain. Suppose the monomial bulb lights up when a certain set of gates take their prescribed values. So, the truncated version of this bulb lights up exactly when these prescribed values are taken by those of these gates that are not included in the considered neighborhood of $u_1$: we simply throw out the binary constraints on the gates of this neighborhood.

It is easy to see that the change in the value of the monomial in our case (a subcase of the second case) is equal to the product of the values of the truncated bulbs, taken with the same coefficient as the monomial itself, multiplied by 1 or -1, depending on case in which the original monomial can be non-zero: in the case when $u_1 = 1$ or when $u_1 = 0$.

Note that the truncated version of the bulb covering $u_1$ is trivial (does not depend on any gates and is identically equal to one). Thus, the number of nontrivial truncated bulbs for a monomial of the type under consideration of degree $l$ does not exceed $l-1$.

Summing up the change in the values (when $u_1$ changes from zero to one) of all monomials, we obtain the required lamp polynomial, bulbs of which depend only on gates outside of neighborhood, of degree at most $d-1$.

The next idea is that in order to prove the cocyclicity of a polynomial, it suffices to verify that for each gate $u_1$, the polynomial of degree at most $d-1$ constructed the way described above, for each possible set of values of the gates of the neighborhood, first, is cocyclic, and secondly, its value is always zero. But we can check the cocyclicity of this polynomial recursively, and it is enough to check the second condition for any set of values of gates outside the neighborhood. Let's do it. The recursion depth will thus be $d$. For constant values of $k$, $d$, the algorithm runs for polynomial time.

A small offtopic about the designation. The name "cocyclic polynomial"\ was chosen because of the topological interpretation of the proof system "cage"\ . I wrote that a cage (presumably) can be associated with a manifold in the corresponding space. During constructing of this manifold, a certain $CW$-complex was constructed and the manifold simply originated from the cocycle of this $CW$-complex of the corresponding dimension. That is, a cocyclic polynomial of degree 1 is a cage, and a cage is a cocycle.

\subsection {Reduction of a proof system "generalized cell"\ to a proof system "cocyclic polynomial"\ .}

Consider a circuit with a "multistorey cage"\ (this is a generalized cage, in which we, however, do not add additional gates to the original circuit, that is, it is just a tree of cages). To the more general case of a "generalized cage"\ (with additional gates), our arguments will be generalized in an obvious way.

Let me remind you that our multistorey cage is a tree, each node of which is a cage, however, defined on the set of choices of gate values, in which the values of some gates are constant.

The first thing we will do is "normalizing each node". We know that the cage corresponding to each node, whenever the gates, which should be constant, take the desired values, has a constant sum of lit bulbs equal to some positive number. We will divide all the weights of the bulbs in a given cage by this positive number. Now the sum of the lit bulbs will always be equal to one. We will perform this operation for each node of the tree.

Let us describe all the monomials of the cocyclic polynomial that we want to associate with a given tree of cages. Now I will describe the process of building such a monomial. Let's choose an arbitrary bulb of the root cage. If its weight is not positive, we will stop there. If its weight is positive, but this bulb implies a contradiction, we will also stop there. If its weight is positive and it does not imply a contradiction, the next level cage should be suspended from it. Let's choose an arbitrary light bulb of this cage. If its weight is not positive, we will stop. If the weight is positive, but the bulb is terminal (implies a contradiction), we stop. If the weight is positive and the bulb is not terminal, the next level cage is suspended from it. Let's choose its arbitrary light bulb. And so on, until we reach a light bulb of non-positive weight or a terminal light bulb. The result is a chain of light bulbs.

We will associate each such chain with a monomial equal to the product of the variables corresponding to the light bulbs included in the chain. We will choose the coefficient of this monomial in our polynomial equal to the product of the weights of the bulbs included in this chain (we mean the weight of the bulb in the cage of the corresponding node after normalization).

Let's check the security condition. Note that the sign of the coefficient at the monomial is determined by the sign of the weight of the last bulb in the corresponding chain (the weights of the remaining bulbs in the chain are positive). If the weight of the last light bulb is positive, then it is terminal, which means that it implies a contradiction. If the weight of the last bulb is not positive, then the monomial coefficient is also not positive, which means that there is no need to look for a light bulb implying a contradiction.

Let's check the cocyclicity. We have described the chains of light bulbs of the cages of the tree, from which we obtain the monomials of the polynomial. Now we are interested in "pre-chains"\ - the initial segments of such chains. By the restriction of a polynomial to a pre-chain, we mean our polynomial, in which we have left only monomials (with the same coefficients), the chains corresponding to which begin precisely with this prechain, and the remaining monomials are simply discarded.

The co-cyclicity of the polynomial follows from the statement that if the gates on which at least one pre-chain bulb depends are set to those values at which all pre-chain light bulbs light up, then the restriction of our polynomial to this pre-chain always takes a value equal to the product of the weights of the light bulbs (as a part of corresponding cages - tree nodes) included in the prechain. If at least one of the gates is set to a different value (and, thus, at least one bulb of the pre-chain does not light up), then the restriction of our polynomial to this pre-chain always takes zero value (the latter is obvious, since in this case all monomials of the restriction are equal to zero).

The first (less obvious) part of the statement is proved by induction on the prechain length from top to bottom. Base. If a pre-chain ends with a terminal bulb or a bulb of negative weight and, thus, is a complete chain, then the statement is obvious for it, since the restriction of the polynomial on it consists of only one monomial with a coefficient exactly equal to the specified product. Transition. If the prechain allows continuation, then all these continuations (that is, chains starting with this prechain) are grouped according to the next bulb of this continuation; moreover, each of the obtained groups, obviously, is a pre-chain of length 1 more; it remains for us to use the induction assumption for each of these prechains and the invariance (and equality to one) of the sum of the weights of the cage of a tree corresponding to these prechains. The statement is proven.

The identity of our polynomial to one follows from the application of the statement to an empty prechain (we assume that the product of the empty set of factors is equal to one).

\subsection{Completeness}.

Here I show the existence of a protected cocyclic polynomial for every unsatisfiable circuit that comes from the work of TM. However, its degree will not be constant, but only polynomial in terms of size of the circuit. When a polynomial has an exponential number of monomials, it complicates practical work with it, but it seems to me that it is still pleasant to realize the completeness of the proof system with which we are working. I hope that in majority of the unsatisfiable examples that come from practice, there exists a constant degree polynomial.

Let's call a light bulb basic if it depends on some gate not of the lower level, as well as on all gates on which this gate in the circuit depends and only on them (the values of the listed gates at which the light lights up are not important, it is only important that it depends exactly on a set of gates of this kind). Consider an arbitrary choice of gate values, as always, with the restriction that the gate, which is considered to be the output given by the circuit, must be equal to 1 (let's call this gate the upper gate). Consider the product of the variables corresponding to the base light bulbs that light up with this choice of gate values. Our polynomial is the sum of such products over all such choices of gate values.

Note that such a polynomial is cocyclic (since for any choice of gate values in which the upper gate is 1, one described product is 1 - the one that corresponds to this choice of values - and the rest are equal to 0, thus the sum is always 1). The security of this polynomial follows from the fact that in any described monomial product there is a light bulb that implies a contradiction (this follows from the impossibility for the circuit to output 1: for any choice of gate values, at which the upper gate is 1, there is a contradiction somewhere).

\subsection {Efficient search for protected cocyclic polynomials of bounded degree.}

Obviously, in the linear space of all lamp polynomials of degree at most $d$, the space of cocyclic polynomials forms a linear subspace. Now I will describe how we can efficiently search for this subspace (in polynomial time).

In the section on local checking of a lamp polynomial for cocyclicity, we discussed lamp polynomials of degree 1 less than the original, which are equal to the change in the value of the original polynomial if we change the value of one of the gates from zero to one, for fixed gate values of the neighborhood of this gate of radius $k$ (we will call such a polynomial a difference polynomial). Note that for a fixed gate (which we change) and its neighborhood, the operation of obtaining the described difference polynomial from the original polynomial is realized by a specific linear operator known to us (from the space of lamp polynomials of degree at most $d$ to the space of lamp polynomials of degree at most $d-1$). Thus, we have a set of linear operators of cardinality linear in the size of the circuit. It is easy to see that for the original polynomial, to be cocyclic, it is necessary and sufficient that each of these linear operators send our polynomial to the space of cocyclic polynomials of degree at most $d-1$, such that always output 0.

Thus, if we have in our hands the space of cocyclic polynomials of degree at most $d-1$, and those that always output 0, then we obtain the space of cocyclic polynomials of degree $d$ by solving a system of linear equations polynomial in size (each linear operator from a list polynomial in size must send a vector, claiming to be a cocyclic polynomial, to the described linear space).

We can recursively find the space of cocyclic polynomials of degree at most $d-1$. It remains to select in it the subspace of such polynomials that always output 0. This can be done by noticing that the value that each of these cocyclic polynomials produces (of course, constant, for each of these polynomials) is a linear function on the space of cocyclic polynomials of degree at most $d-1$, which is easy to find because we can easily compute the value given by each cocyclic polynomial. Thus, it suffices to find such a function and then find its kernel, which is a subspace of codimension 1.

So, now we can find the space of cocyclic polynomials of degree at most $d$ in polynomial time. Let us now show how we can look for a protected cocyclic polynomial of degree at most $d$ or verify its absence in polinomial time. To do this, it suffices to note that a protected cocyclic polynomial is basically a cocyclic polynomial that satisfies a polynomial set of linear inequalities: each such inequality is a condition that the coefficient in some monomials must not be positive. We can simply go through all the products of no more than $d$ variables corresponding to small bulbs and see if a given set of bulbs contains a bulb that implies a contradiction. If it does not exist (we will call such monomials unprotected, if such a light bulb still exists - we will call it protected), the coefficient of this product should not be positive. Systems of a polynomial number of linear equations and inequalities over the field of rational numbers can be solved in polynomial time, as was proved, for example, in ~\cite {LP}.

I have an assumption that each cocyclic polynomial can be represented as a linear combination of products of several (in the number not exceeding the degree of the polynomial) cocyclic polynomials of degree 1. If so, then this will allow finding the space of cocyclic polynomials of fixed degree in a slightly simpler way. However, I have not yet verified this assumption.

\section {General algorithm}

Suppose that we, based on the resources available to us, chose a sufficiently large constant $d$ and tried to find a protected cocyclic polynomial of degree at most $d$ for the circuit, but we failed to do this and found out that there is no such polynomial. Then it's time to try to find an arrangement of values in the gates of the circuit that does not violate any gate (in which each gate takes exactly proper value, according to the gates on which it depends in the circuit), and input line to the circuit as well, which is the answer to the task. This section explains how to find such an arrangement of values.

The idea is that other cocyclic polynomials (not necessarily protected) can help us with this. Imagine, for example, a situation where we have found a cocyclic polynomial in which there is only one unprotected monomial, the coefficient of which is positive, and the polynomial itself always takes some positive value. This means that if we want to arrange values into gates that do not violate any gate, then all the bulbs corresponding to this monome should light up in it. Because since there are no violated gates, in each protected gate there is at least one light bulb (implying a contradiction) that does not light up, and therefore each protected monom is reset to zero; and where to get positive terms, if not from unprotected monomials with a positive coefficient?. And this information (about the light bulbs that should light up) can be used to build the desired arrangement of values.

In the same way, not only positive (as I call polynomials that always take a positive value) cocyclic polynomials with a single positive unprotected monomial can be useful, but also, more generally, positive cocyclic polynomials, in which there are little positive unprotected monomials. For example, if a polynomial of degree 1 always takes the value 1 and its coefficients for unprotected monomials of degree 1 (that is, in fact, for small bulbs) are $k$ ones, $10k$ minus ones and maybe some number of zeros, and the free coefficient of this polynomial is zero, then this is, let's say, an argument in the direction of lighting up for unprotected bulbs, the coefficient at which is 1 and an argument for not lighting up for unprotected bulbs, the coefficient for which is -1 (among the latter, no more than one tenth can light up, and exactly one less bulb should light up among them than among the former). Of course, the same is true for large degree cocyclic polynomials.

And we can collect many different cocyclic polynomials, each of which gives a certain number of such arguments of different strengths, for different bulbs, indicating that the corresponding bulb should light up or should not light up. And if for some light bulb the total strength of the arguments in the direction of lighting up turns out to be much greater than the total strength of the arguments in the direction of not lighting up, then you should think about choosing the gate values so that the corresponding light bulb lights up.

Now let's talk about the algorithm more detailed.

We will pursue to obtain the desired arrangement of values in gates from the final state of one structure modified in time. The structure is following. Consider a certain system of small bulbs, as always, each lights up (takes the value 1) if and only if the gates on which it depends take a strictly defined set of values (otherwise it does not light up, taking the value 0). The structure is a set of numbers $p_i$, $0 \le p_i \le 1$, ($i$ ranges from 1 to the number of bulbs) such that if the index $i$ corresponds to a bulb implying a contradiction, then $p_i = 0$. We will call this structure the light bulb probability distribution.

Let's call an arbitrary mapping of zeros and ones to the bulbs of our system of bulbs a situation. That is, a situation is an act of instructing some bulbs to be lit, and the rest not to be lit, and it is not at all necessary for the situation to come from some arrangement of zeros and ones in the gates (hereinafter we will call them simply arrangements). We will be interested in the situation as a random variable in which the $i$ -th bulb is lit with the probability $p_i$ and whether the bulb is lit or not is selected for each bulb independently.

The algorithm will change the values of the probabilities $p_i$ so that the probabilities corresponding to the bulbs implying a problem are always zero. As a result, we want to see the probability distribution on the light bulbs, in which each probability is either zero or one. Thus, it is assumed that in the final position, our random variable produces exactly one fixed situation. Another thing that we want to see in the end is that the only situation that our random variable always produces has the property that it comes from some arrangement of zeros and ones in the gates. That is, there is such an arrangement that the light in it lights up if and only if it lights up in our "final" situation (hereinafter I will call it the final situation - without quotes). We will call a situation with this property consistent.

The cocyclic polynomial can be interpreted in an obvious way as a function of the arrangement and as a function of the situation. The following statement gives a criterion for the consistency of the situation, under some condition for the system of light bulbs.

Lemma. If a system of light bulbs satisfies the property that, together with a bulb depending exactly on some set of gates $A$, it contains all bulbs satisfying the property that they depend, exactly, on the set of gates, which is some subset of $A$ (there is a finite (and constant) number of such bulbs for each $A$: two bulbs that always light up or do not light up at the same time, we equate) - we will call this the inclusion closeness property - then the situation in this system of light bulbs is consistent if and only if any cocyclic polynomial of degree 1 outputs in this situation the same value that it gives in any arrangement.

Proof. The "only if"\ part is obvious. Let us prove that if each linear cocyclic polynomial outputs on a situation exactly the same value that it outputs on all arrangements (or, which is the same, on all situations that come from the arrangements), then the situation is consistent.

Let some light bulb depend on some set of gates $A$ (we will call this set $A$ basic). Thus, by the inclusion closeness property, our system of bulbs contains all bulbs, each of which satisfies the property that the set of gates on which this bulb depends coincides with $A$. Consider a linear cocyclic polynomial in which all these bulbs are included with a coefficient of 1, and all other bulbs - with a coefficient of 0 (the free coefficient is 0). Obviously, it is cocyclic and outputs 1 in all arrangements. This means that in our situation it should also output 1, which means that exactly one of the considered bulbs should light up in our situation, and the rest should not light up.

For the basic set $A$, thus, it is possible to uniquely determine the set of values for the gates of this set, to which in our situation the value 1 is assigned. We will call this set of values the realization of the set $A$ in this situation. It is easy to see that the consistency of the situation now follows from the statement that, for any two base sets $A$ and $B$, their realizations are concerted at their intersections (if we prove this statement, then it will be enough to choose such an arrangement whose restriction to each basic set coincides with the realization of this set). Let us prove this statement. By the inclusion closeness property, in our system of light bulbs, in addition to $A$ and $B$, there is a set $A \cap B$, so it suffices to prove the concertedness at the intersection of the realizations $A$ and $A \cap B$, as well as, concertedness at the intersection of the realizations $B$ and $A \cap B$. By virtue of the symmetry of these statements, it is enough to prove the first.

Consider a cocyclic polynomial, with coefficient 1 for bulbs, each of which has the property that its base set coincides with $A$ and that it lights up when the gates take strictly defined set of values, different for each such bulb and it runs through all possible such sets, with the restriction that on $A \cap B$ this set of values coincides with the realization of $A \cap B$; and also, with a coefficient -1 for a light bulb with a basic set $A \cap B$, which lights up when a set of gate values from $A \cap B$ coincides with the realization of $A \cap B$; the other coefficients of the polynomial are equal to 0.

It is easy to see that the polynomial is indeed cocyclic and outputs 0 on all arrangements. It is also easy to see that if the realizations of $A$ and $A \cap B$ are not concerted on $A \cap B$, then the polynomial in our situation will return -1. Contradiction. So, the realizations of the basic sets are concerted at their intersection. The lemma is proved.

So, if all the values of the bulbs in the final situation are good in the sense of linear cocyclic polynomials, then they give an arrangement that does not violate any gate.

We will work under the assumption that our system of light bulbs is closed by inclusion, for example, it can be a system of light bulbs with a diameter not exceeding $k$, for a constant $k$. Although many systems that are not closed by inclusion would fit.

Let's go back to the algorithm. So, we want to know how to change the probabilities of light bulbs in such a way that in the end each of them equals to zero or one, and in the resulting situation obtained in this way, all the cocyclic polynomials output the correct value (we could limit ourselves only to linear cocyclic polynomials, but our algorithm will use large degree polynomials).

At each moment of time we will maintain a set of cocyclic polynomials $\Omega$, which we will supplement from time to time and, possibly, discard polynomials from it.

For each light bulb $s$ from our system of bulbs and a cocyclic polynomial $f$ from the current set, we will evaluate 2 numerical values, one corresponds to the position of the light bulb $s$ in the off state (let's call it $T_0 = T_0(s, f)$) , the other - to its position in the on state (let's call it $T_1 = T_1 (s, f)$).

So, the on position of the $s$ light bulb. Consider such a probability distribution on situations $K_j = K_j (s)$. For each $i$, if the $i$-th bulb does not coincide with the $s$-th bulb, we randomly and independently of the other bulbs choose the state of the $i$-th bulb: with the probability $p_i$ we make it on and with the probability $1 - p_i$ we make it off ($p_i$ is the current value of the probability at the $i$-th bulb). We make the bulb $s$ lit on if $j = 1$ and lit off if $j = 0$. We will be interested in the random variable $F_j$, which is calculated as the value of the polynomial $f$ in a random situation from the distribution $K_j$.

Recall that each cocyclic polynomial $f$ has a corresponding value of $t = t(f)$, which is equal to the value that $f$ takes in all consistent situations. So, the value $T_j$ we need is calculated as minimum of two probabilities: the probability that $F_j \le t(f)$ and the probability that $F_j \ge t(f)$.

We also need the value $T = T(f)$, which is calculated in a very similar way, differing only in that we do not select any light bulb $s$: we build a random situation where each light bulb turns on or off with the current assigned probability $p_i$ (we will call such a distribution on situations $K$), calculate the value of $f$ on it and choose the minimum of two probabilities: the probability that the resulting random variable is not greater than $t(f)$ and the probability that the resulting random variable is not less than $t(f)$.

Let's talk about how we can evaluate $T_0$, $T_1$. Consider the distribution on situations $S_i$ We have a real-valued random variable (equal to a given function of a random situation), which can take an exponential number of values in the size of the circuit, and we want to estimate how often it takes a value, say, not exceeding a certain number. We will evaluate simply by generating a random situation from a given distribution $K_j$ a sufficiently large number of times, polynomial in size $N$ of the circuit (say, $p(N)$, where $p$ is some polynomial) (we can, independently of the other light bulbs make the $i$-th light bulb lit with probability $p_i$ using a pseudorandom number generator) and calculating the value of $f$ in this situation. Doing this we get a polynomial set of numbers.

Maybe some of these numbers will be less than $t(f)$, some will be greater, and some will be equal to $t(f)$. Let's calculate two numbers: $v_{j.0}$, equal to the number of those of these numbers that do not exceed $t(f)$ and $v_{j.1}$, equal to the number of those of these numbers that are not less then $t(f)$. In case both $v_{j.0}$ and $v_{j.1}$ are greater than zero, for both values of $j$ (first opportunity), we estimate $T_j$ by minimum over $i$ of $v_{j.i}/p(N)$.

If exactly one of $v_{j.i}$ (over all values of $i$, $j$) is equal to zero (second possibility), we estimate $T_j$ by minimum over $i$ of value $v_{j.i}/p(N)$, if this minimum is nonzero; if this minimum is equal to zero, we estimate $T_j$ as $1/2p(n)$.

If, for each $j$, there exists an $i$ such that $v_{j.i} = 0$ (the third possibility), our estimate will be a little more complicated and somewhat more rough. It would be possible to estimate both values of $T_j$ as $1/2p(n)$, but the point is that we will be interested in how $T_0$ and $T_1$ correspond to each other, so we want the relation between the estimates of these quantities to even remotely approximate the ratio of the values themselves. Therefore, we will do this. Let's choose a not very large, although not small positive constant $h$. We have two polynomial in size sets of values that were taken by our random variables $F_0$ and $F_1$. Let's place them on separate number lines. Next, we will walk on each of these lines from the point $t(f)$ in the direction where all the points of the corresponding set of values lie, with the same constant speed and we will walk until we collect at least $h$ points on each of the lines. Suppose we have collected $a_0$ points on the line corresponding to index 0 and $a_1$ points on the line corresponding to index 1. Then our estimate of $T_j$ will be $a_j/((a_1 + a_2)p(N))$. Again, this estimate is rather rough, but it has a pretty good chance of at least "guessing"\ which of $T_j$ is greater and "catching"\ the case when $T_j$ are very different (in this case, these estimates also have a good chance to differ greatly in absolute value).

Before continuing, I will introduce a real-valued function $U = U (s, f) = R (T_0 (s, f), T_1 (s, f))$. The $U$ function represents the very "power of argument"\ that I wrote about closer to the beginning of this section. I'll define $ U $ by defining $ R $.

$ R (u, v) = l (u / (u + v)) m (1 / (u ^ 2 + v ^ 2)) $, where $ m $ is some increasing positive-valued continuous function, and $l$ is function on the interval $(0,1)$ satisfying the properties that, firstly, $ l (1/2) = 0 $, and, secondly, $ l (1/2 + t) = -l (1 / 2 - t) $, for any $ t $, $ 0 <t <1/2 $ and thirdly, on the interval $ (1/2, 1) $, $ l $ is an increasing positive-valued function, and we require that $l$ tends to infinity when approaching 1.

Let's pay attention to how the function $ U $ behaves. If it is greater than zero, then $ T_1 (s, f) <T_0 (s, f) $. If it is less than zero, then $ T_0 (s, f) <T_1 (s, f) $. If it is equal to zero, then $ T_1 (s, f) = T_0 (s, f) $. Note also that $ U $ tends to increase in absolute value when $ T_0 (s, f) $ and $ T_1 (s, f) $ decrease in absolute value (due to the multiplier with the function $ m $).

Let's call a polynomial $ f $ discriminated if $ T (f) $ is small. It would be more correct to talk about the degree of discrimination of a polynomial: the smaller the value of $ T $ for it, the more the polynomial is discriminated. But for the sake of brevity, we will refer to heavily discriminated polynomials as simply discriminated. The polynomial $ f $ is discriminated if, in a random situation from the distribution $ K $, the polynomial $ f $ calculated in this situation gives a value, IN MOST CASES, greater than $ t (f) $, or, in most cases, less than $ t (f) $. And the more significant this majority is, the more discriminated the polynomial is.

We will change the values of the probabilities $ p_i $, mainly in such a way as to make the polynomials from the list $ \Omega $ as less discriminated as possible. Moreover, we are much more interested in reducing the extent of discrimination of highly discriminated polynomials than their less discriminated counterparts.

Consider, for example, some polynomial $ f $ from $ \Omega $ and some light bulb $ s $. Suppose that $ T_1 (s, f) <T_0 (s, f) $, and, as well as for $ j = 0 $ and $ j = 1 $, the probability that the value of $ f $ on a random situations from the distribution $ K_j $ is more than $ t (f) $, is more than $ 1/2 $ (or less than $ 1/2 $ for both values of $ j $). Note that by decreasing $ p_s $ we decrease the extent of discrimination of $ f $. Indeed, if we change $ p_s $ linearly (in one direction, with the constant speed), then the probability that, in a random situation from the distribution $ K $, $ f $ will output a value less (or greater) than $ f ( t) $, will change linearly, since the probability of each situation will change linearly. Moreover, when the value of $ p_s $ is 1, this probability will be equal to $ T_1 (s, f) $, and when the value of $ p_s $ is equal to 0, this probability will be equal to $ T_0 (s, f) $.

Under similar conditions, with the only difference that $ T_1 (s, f)> T_0 (s, f) $, in order to decrease the extent of discrimination of $ f $, we need to increase $ p_s $.

If, say, $ T_1 (s, f) <T_0 (s, f) $, and also, the probability that the value of $ f $ on a random situation from the distribution $ K_0 $ is greater than $ t (f) $, is greater than $ 1/2 $, and the probability that the value of $ f $ in a random situation from the distribution $ K_1 $ is greater than $ t (f) $, is less than $ 1/2 $, then with a linear decrease in $ p_s $, the extent of discrimination may first decrease, and then start to increase, nevertheless, at $ p_s = 0 $, $ f $ is less discriminated than at $ p_s = 1 $.

I wrote that the value of $ U (s, f) $ represents the power of the argument for changing $ p_s $ in one direction or another. So, let us clarify that if $ U (s, f)> 0 $, then from what we have just written, we see that it makes sense to decrease $ p_s $, and if $ U (s, f) <0 $, then it makes sense to increase $ p_s $. And the more $ U (s, f) $, modulo, the more sense it makes to change $ p_s $ in this way.

For example, if the polynomial $ f $ is to a large extent discriminated and in random situations, as a rule (with probability $ 1 - \varepsilon $), takes a value less than $ t (f) $, and for $ s = 1 $, the probability that $ f $ will take a value greater than or equal to $ t (f) $ is slightly greater than the same probability at $ s = 0 $, then in order to reduce the extent of discrimination of $ f $, it makes sense to increase the probability that $ s = 1 $. Moreover, I wrote that the less $ \varepsilon $ (ie, the more discriminated the polynomial), the more we want to "get him out of trouble" \, that is, reduce its extent of discrimination, but exactly for small values of $ \varepsilon $, $ U (s, f) $ tends to take the largest values modulo (for this we introduced a multiplier with the $ m $ function when defining $ U $).

At every regular moment of the algorithm's work (there will be irregular moments of its work, the word "regular"\ in this context is close in meaning to the word "normal" \, we endow it with the same meaning as it is usually given in mathematical texts), for for each light bulb $ s $ we will find the total argument $ L (s) $ for the change of probability $ p_s $. It is simply the sum over all $ f \in \Omega $ of $ U (s, f) $. If the total argument turns out to be greater than zero, we will decrease $ p_s $ by a small amount (naturally, depending on $ L (s) $), if it is less than zero, we will increase it, if equal to zero, we will not change it. We keep the magnitude of the change in $ p_s $ small because we want to approximate the continuous change in probabilities.

How does the magnitude of change depend on $ L (s) $? Let's restrict ourselves to changing the probability $ p_i $ so that the value $ \tan ({\frac{\pi}{2}} (2p_i - 1)) $ changes exactly by $ - \lambda L (i) $, where $ \lambda $ is a small positive constant, which we choose as small as we want to bring the change in probabilities closer to continuous.

The arguments corresponding to the most discriminated polynomials are the largest in absolute value and it is assumed that they often make a decisive contribution to the total argument.

Let us now talk about why we reduce the extent of discrimination of cocyclic polynomials. The point is that $ T (f) $ correlates quite well with the probability that, in a random situation from the current distribution $ K $, the value of the polynomial $ f $ IS EQUAL TO $ t (f) $.

Let me explain. Suppose we are dealing with a linear cocyclic polynomial $ f $. Then, by the central limit theorem, the distribution of values of $ f $  on situations from the distribution $ K $ is close to normal (we take the sum of independent random variables corresponding to the light bulbs: if the $ i $-th light bulb lights up (and this happens with the probability $ p_i $ ), such a random variable is equal to the coefficient in $ f $ for this light bulb, if the light does not light up, such a random variable is zero). That is, the density of such a distribution is close to a Gaussian on a straight line.

And if $ f $ is discriminated, it means that the mark $ t (f) $ is far from the median point of this Gaussian (as I call the point at which the Gaussian reaches its maximum) and that part of the subgraph of this Gaussian that is on one side of $ t (f) $ (the smaller of these two parts) is a small part of the area of this subgraph (this part roughly coincides with $ T (f) $). That is, $ t (f) $ kind of bites off a rather small piece of the Gaussian. If $ f $ is not discriminated, $ t (f) $ hits closer to the median point of the Gaussian and breaks it down into parts that are more comparable in size. It is clear that in the first case the density of our distribution at the point $ t (f) $ (that is, the very value of the Gaussian at this point) will be slightly less than its density in the second case. And this means that in the first case, in a random situation, we will get a value exactly equal to $ t (f) $, as expected, with a lower probability than in the second.

In the case of polynomials of a greater degree, I do not yet know an explicit class of functions that approximate the distribution of the values of such a polynomial on random situations from the distribution of $ K $, as in the case with linear polynomials and Gaussians. It is possible that the explicit form of such functions can be specified (and this, by the way, would allow us to more accurately estimate $ T (f) $, $ T_0 (f) $ and $ T_1 (f) $, in the case that one of these values turned out to be small and on one of the sides of $ t (f) $ none of the values of $ f $ from our polynomial sample was found - we would just pick a function from this explicit class of functions that best approximates our distribution and estimated our values as if it were the true distribution density - later I will tell you how to slightly modify the algorithm so that this class of functions is, again, the class of Gaussians, that is, our random variable will be distributed normally - and this, as I said, will allow us to more accurately estimate $ T (f) $, $ T_0 (f) $ and $ T_1 (f) $, when the overwhelming part of the Gaussian subgraph is on one side of $ T (f) $). But if not, we will still allow ourselves to talk about the distribution density at a given point, although I do not give an explicit way to unambiguously compare our discrete distribution of values of $ f $ in random situations from $ K $ with a continuous distribution density function approximating this discrete distribution ...

Most often, the density of our distribution, while walking along the number line to the left (or to the right), will not decrease abruptly - most likely it will decrease smoothly, gradually tapering off (it is unlikely that at a certain moment there will be a sharp break, after which the density will become zero). This means that in this case, as in the case of the Gaussian, if we take a polynomial $ f $ with very small $ T (f) $, the density of the corresponding distribution at the point $ t (f) $ turns out to be expectedly less than a similar value for the polynomial $ g $ with $ T (g) $ close to $ 1/2 $.

So, we look at $ T (f) $ as at an estimate of the probability that in a random situation from the current distribution $ K $, the value of the polynomial $ f $ is equal to the "correct"\ value, that is, $ t (f) $. In exactly the same way, one can look at $ T_j (s, f) $ as at an estimate of the probability that, in a random situation from the current distribution $ K_j (s) $, the value of the polynomial $ f $ is equal to $ t (f) $ (in the sense that from the value of $ T_j (s, f) $ you can extract a number that is expected to approximate the given probability, and this number is the smaller, the less is $ T_j (s, f) $).

It is possible that in order to estimate these probabilities, it would be better to work not with the functionals $ T (f) $, but to estimate the density of the distribution at the point $ t (f) $, observing how often the values of $ f $ on random situations from $ K $ fall into a small neighborhood $ t (f) $, but for some reason I decided that it is better to deal with functionals $ T (f) $.

As I have already mentioned, the work of our algorithm at regular moments of time will consist of changing the probabilities $ p_i $. A specific way to change these probabilities has been described above. Changing these probabilities is aimed at making the polynomials from the list $ \Omega $, and in particular, the most discriminated polynomials from this list, as little as possible.

But we have seen that the value $ T (f) $, which determines the extent of discrimination of the polynomial $ f $, correlates well with the probability that the value of $ f $ in a random situation from the current distribution $ K $ will take the value $ t (f) $. Therefore, our process of changing probabilities can be viewed as a process that "continuously"\ changes probabilities in such a way that for each polynomial $ f $ from $ \Omega $ a randomly chosen situation from the distribution $ K $ (that is, $ i $-th light bulb turns on with probability $ p_i $, independently of other light bulbs) with the highest possible probability had the property that $ f $ on it takes the value $ t (f) $. This means that we jointly optimize this probability for all polynomials from $ \Omega $, finding some compromise.

Our algorithm tries to increase this probability for each polynomial $ f $ from $ \Omega $, and especially for those of these polynomials for which such a probability is small (of course, not the real probability - we do not know it - but our estimate of this real probability).

Recall, for example, the cocyclic polynomial $ f $ of degree 1 described closer to the beginning of the section, which always takes the value 1 and its coefficients for unprotected monomials of degree 1 (that is, in fact, for small light bulbs) are $ k $ ones ( we call such bulbs as $ A $ bulbs), $ 10k $ minus ones ($ B $ bulbs), and maybe some number of zeros, and the free coefficient of this polynomial is zero. If initially all probabilities tied with unprotected bulbs (not implying a contradiction) are equal to $ 1/2 $ (remember, probabilities with protected bulbs are always zero - protected bulbs are always off), then initially we have an argument towards moving the probabilities for $ A $ bulbs from $ 1/2 $ towards one: when a particular $ A $ light bulb is on, the probability that $ f $ will take the value $ t (f) $ is initially slightly higher than when it is off. However, over time, the situation may change: after the algorithm has worked for a certain time, for example, the probabilities for $ B $ bulbs can become, for the most part, very close to zero, but with $ A $ bulbs - close to 1 And then, for a particular $ A $ light bulb, $ f $ will generate an argument in the direction of moving corresponding probability towards zero.

Now I will give the final algorithm for finding a situation, and then I will motivate its details.

The algorithm will use the procedure for finding and adding new cocyclic polynomials to the list of processed cocyclic polynomials $ \Omega $. This procedure will be the subject of the next section, in which the procedure will be described and detailed. For now, I will use it as a black box, and I will only say that the procedure will try to find the most discriminated and semi-discriminated (I will give the definition of the latter later) cocyclic polynomials for the current values of probabilities.

Also, the algorithm will use the operations of fixing and unfixing of unprotected bulbs. Fixing the $ i $ -th light bulb means freezing the value of $ p_i $ in state 0 or 1 (just before the moment of freezing, the probability value can differ from 0 and 1). As long as the light bulb is fixed, we do not change the probability value for this light bulb. Unfixing the $ i $-th light bulb means unfreezing the value of $ p_i $, allowing it to change (immediately after the moment of unfreezing, the probability value is exactly the same as it was in the frozen state). Thus, at any moment of time, light bulbs are divided into 2 types: fixed, the probability for which is temporarily fixed and takes the value 0 or 1, and non-fixed, the value of which varies within the segment $ [0,1] $.

Also, at any given time, there will be some subset of bulbs in the "unfixable" status among the unfixed bulbs. If the light bulb is in this state, it signals to the algorithm that it cannot be fixed. The rest of the unfixed light bulbs are in the "fixable" status.

At the moment, the algorithm is a little unfinished, in the sense that it uses a number of parameters (a set of numbers and two real-valued functions of a real domain), the most optimal value of which is still unknown to me. We only know the magnitude of these parameters (if we are talking about numbers) and how they should roughly relate to each other and to the size of the circuit. It might be worth experimenting to find the most appropriate values for these parameters. The efficiency of the algorithm is unlikely to be sensitive to slight changes in these parameters.

For brevity, I will call these parameters selectable.

When describing the algorithm, for the sake of convenience, I call the selecatble parameters-numbers constants, but in reality they are not really constants, but numbers, possibly depending on the size of the circuit.

We have already introduced some of these parameters above.

The list of parameters is as follows: $ d $ (the degree of the cocyclic polynomials we are working with), $ a $ (the length of the time interval after which we unfix the light bulb from the beginning of the list), $ b $ (the length of the time interval after which we update the list of cocyclic polynomials), $ m_1 $ (the length of the first step inside the cycle), $ m $ (the maximum length of the second step inside the cycle), $ C $ (the time during which the bulbs remain unfixed), $ L_1 $ (the threshold for unfixing a fixed bulb), $ \delta $ (the size of the neighborhoods of zero and one, that we fix the light bulb, the probability hits within this neighborhood), $ p (N) $ (the size of the sample with which we estimate $ T_j $), $ h $ (the parameter used to estimate $ T_j $), $ l (t) $ and $ m (t) $ (real-valued functions used to define U (s, f)), $ \lambda $ (small real-valued multiplicative constant, via which we reduce the change in probabilities by each step, bringing the change in probabilities closer to continuous).

So, the algorithm. Comments to it are shown in curly braces.

\par \bigskip
\par \bigskip
\par \bigskip

* Check if there are no protected cocyclic polynomials of degree at most $ d $, if such a polynomial is found, stop the algorithm and report that the situation does not exist

* $ r $: = 0;

$ \{$ $ r $ is the counter we need in order to understand how long the algorithm is running, in order to update the list of cocyclic polynomials from time to time, to unfix the light bulb from the beginning of the list, which will be discussed later, as well as, in order to keep track of how long each light bulb in the "unfixable"\ status stays in this status, if you count from the moment when it acquired this status the last time. $ \} $

* $ p_i $: = 0.5, for all unprotected $ i $

$ \{$ we initialized the initial probabilities for unprotected bulbs $ \} $

* $ p_i $: = 0, for all protected $ i $

$ \{$ you can forget about protected bulbs altogether, they are always off and the probabilities are always equal to zero $ \} $

* initialize $ P $ list as empty

$ \{$ At any given time, the $ P $ list will contain all currently fixed bulbs, in the order in which they were fixed last time; thus, at the beginning of the list, there will be a light bulb that is in a fixed state, since the moment of the last fastening, the longest time among all fixed bulbs. $ \} $

* Firstly all unprotected bulbs are unfixed, we initialize their status as "fixable"\ ;

* As long as there is at least one unfixed value of an unprotected light bulb, or all of them are fixed, but the corresponding situation is not consistent, we do:

$ \{$ The main part of the algorithm's work consists of one cycle, the body of which consists of three steps. The first of these steps is optional: I added it because my intuition (which I cannot explain yet) tells me that the algorithm will work better if there is a substantial period of time between two adjacent fixings of light bulbs (this first step is exactly aimed to "wait"\ this substantial period of time, simply changing the probabilities and not fixing anything, but possibly unfixing). However, perhaps everything will be fine if the first step is simply removed.

After completing each iteration of the loop, the algorithm checks if we have reached the goal. This check consists in that, firstly, we check if all unprotected bulbs are fixed (if not all, then we have not yet arrived at the desired result). Secondly, if all the bulbs are fixed, it is necessary to check if the corresponding situation is consistent. The corresponding situation is understood as a situation in which unprotected bulbs, in which the probability is 1, are turned on, and unprotected bulbs, in which the probability is 0, are not turned on (protected bulbs, as always, are off).

Whether a particular situation is consistent is checked locally at each point in the circuit. $ \} $

    \hspace {1cm} * (optional): do $ m_1 $ times:
		
$ \{$ This is the first step. It is a loop that repeats $ m_1 $ times, where $ m_1 $ is the selectable parameter. $ \} $

    \hspace {2cm} * For each unprotected light bulb $ i $, we calculate the total argument $ L (i) $ as described above;
        
    \hspace {2cm} * Change all probabilities $ p_i $ corresponding to unprotected unfixed bulbs as described above;
        
    \hspace {2cm} * For each fixed unprotected light bulb $ s $, such that the probability corresponding to it, at the moment, is 1, if the calculated value of the total argument $ L (s) $ with it is greater than the constant $ L_1 $ we have chosen, then we unfix the light bulb $ s $, remove it from the list $ P $ and declare it unfixable. Symmetrically, for each fixed unprotected light bulb $ s $, such that the probability corresponding to it, at the moment, is equal to 0, if the calculated value of the total argument $ L (s) $ for it is less than $ -L_1 $, then we unfix the light bulb $ s $, remove it from the list $ P $, and declare it to be unfixable;
        
    \hspace {2cm} * $ r $: = $ r + 1 $;
        
    \hspace {2cm} * if $ r $ is divisible by $ a $, unfix the light bulb from the beginning of the list $ P $, remove it from the list $ P $ and declare it unfixable;
        
    \hspace {2cm} * if $ r $ is divisible by $ b $, we look for and add several (in the amount of the chosen parameter) new cocyclic polynomials (see the next section);
        
    \hspace {2cm} * if an unprotected light bulb is in the "unfixable"\ status for the time $ C $ (since the moment it changed its status to "unfixable"\ the last time) ($ C $ is a selectable constant; times are measured in values of counter $ r $: if the counter value has changed by 10, 10 units of time have passed), we change its status to "fixable"\;

    \hspace {1cm} * do $ m $ times ($ m $ is the constant we have chosen), or until we had to fix something:
    
$ \{$ This is the second step. It is a cycle, the body of which differs from the body of the cycle of the first step only in that some kind of light bulb is possibly fixed in it. $ \} $

    \hspace {2cm} * For each unprotected light bulb $ i $, we calculate the total argument $ L (i) $ as described above;
        
    \hspace {2cm} * Change all probabilities $ p_i $ corresponding to unprotected unfixed bulbs as described above;
     
    \hspace {2cm} * If any of $ p_i $, for unprotected fixed $ i $, differs from 0 or from 1 by no more than $ \delta $, where $ \delta $ is the constant we have chosen, by less than $ 1 / 2 $, we fix the $ i $-th bulb, making $ p_i $ equal to 0 or 1, respectively (if there are several such bulbs, we do this, for definiteness, with bulb of the smallest number); In this case, as well, we add the light bulb we just fixed to the end of the list $ P $, and also interrupt the execution of the cycle of the second step and go to the third step;
     
    \hspace {2cm} * For each fixed unprotected light bulb $ s $, such that the probability with it, at the moment, is 1, if the calculated value of the total argument $ L (s) $ with it is greater than the chosen by us constant $ L_1 $, then we unfix the light bulb $ s $, remove it from the list $ P $ and declare it unfixable. Symmetrically, for each fixed unprotected light bulb $ s $, such that the probability with it, at the moment, is equal to 0, if the calculated value of the total argument $ L (s) $ for it is less than $ -L_1 $, then we unfix the light bulb $ s $, remove it from the list $ P $, and declare it to be unfixable;
        
    \hspace {2cm} * $ r $: = $ r + 1 $;
        
    \hspace {2cm} * if $ r $ is divisible by $ a $, unfix the light bulb from the beginning of the list $ P $, remove it from the list $ P $ and declare it unfixable;
        
    \hspace {2cm} * if $ r $ is divisible by $ b $, look for and add several (in the amount of the selectable parameter) new cocyclic polynomials (see the next section);
        
    \hspace {2cm} * if an unprotected light bulb is in the "unfixable"\ state for time $ C $ (since the moment it changed its status to "unfixable"\ the last time), we change its status to "fixable"\;

    \hspace {1cm} * if at the previous (second) step we did not have to fix anything (the cycle worked through all $ m $ loops and none of the probabilities of unprotected bulbs never approached neither zero nor one), we fix that unprotected fixable light bulb, the probability value at which is farthest from $ 1/2 $ at the moment; at the same time, if this value is closer to 0, we fix it to 0, and if the value is closer to 1, we fix it to 1 (if there are no unprotected fixable bulbs at all, we do nothing at this step);

$ \{$ This was the third step. $ \} $

\par \bigskip
\par \bigskip
\par \bigskip

First, we will motivate that element of the algorithm, which I call fixing the light bulb. We need fixing in order to get a set of probabilities, each of which is equal to zero or one. If we did not fix the light bulbs, but simply changed the probabilities, we would greatly reduce our chances of getting a set of probabilities, each of which is very close to zero or one, and hence the situation. The change in the probabilities with unprotected unfixed bulbs, in turn, is necessary for us in order to understand which of the bulbs should be fixed next and in what value (0 or 1).

I already wrote above that we change the probabilities in such a way that a random situation from the distribution $ K $ satisfies with the highest probability that the value of $ f $ on it is equal to $ t (f) $, finding a compromise of this probability for polynomials $ f $ from $ \Omega $. Therefore, it is fair to say that we increase the expected probability $ p $ that a random situation from the distribution $ K $ will be consistent. The same remains true in the case when we start from time to time to fix and unfix the light bulbs in the manner described in the algorithm. This probability may perhaps expectedly decrease slightly at the moment when we fix the light bulb.

Imagine a situation where we do not change the probabilities: all probabilities with unfixed bulbs are always equal to $ 1/2 $ and from time to time we randomly choose a bulb and fix it, making the probability 0 or 1 with the same probability. The probability of $ p $, in this case, in the end, most likely, will become equal to 0, since in the end we will obtain the distribution of $ K $, always giving the same completely arbitrary situation. That is, along the way, at a certain moment, we will most likely simply "lose all consistent situations"\ - $ K $ will no longer output them at all and $ p $ will become equal to 0.

What can prevent us from "losing all the consistent situations along the way"\ in our case? First, between two adjacent fixings of light bulbs, we intentionally change the probabilities so that the probability $ p $ is expected to increase. Secondly, for fixing we always choose a light bulb, which at the moment corresponds to the probability farthest from $ 1/2 $ (or one of the most distant, in the case of "premature"\ fixation, when we find a probability that is farther no more than by $ \delta $ from zero or one and there are several such probabilities at once), let it be $ s $ and let it correspond to the probability $ q $. Let, also, for definiteness, $ q> 1/2 $. This means that choosing a random situation from the version of the distribution $ K $ before fixing, we, in fact, with the probability $ q $ choose the situation from the distribution $ K_1 (s) $ and with the probability $ 1 - q $ choose the situation from the distribution $ K_0 (s) $. And choosing a random situation from the version of the distribution $ K $ after fixing, we choose a situation from the distribution $ K_1 (s) $. It is clear that the closer the probability $ q $, in this case, to 1, the less expectedly the probability $ p $ will change as a result of fixation.

For similar reasons, when we choose a light bulb for fixing with the probability farthest from $ 1/2 $, the values of $ T (s, f) $ also change expectedly the weakest.

From time to time, while we change the probabilities and fix the bulbs, we may have serious arguments to unfix a particular bulb. For example, this can happen when we have found (and added to $ \Omega $) a new cocyclic polynomial $ f $, which for sure takes a value less (or greater) than $ t (f) $ in those situations in which the values of those bulbs , which we fixed, correspond to the values of the probabilities fixed with these bulbs: in this case, at least one of the fixed bulbs must definitely must be unfixed. Such arguments can also arise from polynomials added to $ \Omega $ long ago, as a result of changing probabilities. We unfix the light bulbs "reluctantly"\: we do this only if the arguments to do so are strong enough, to be more precise, these arguments in total must be greater than the selectable constant $ L_1 $.

We also sometimes unfix a light bulb from the beginning of the list $ P $, even if there are no serious arguments to unfix it, which were discussed above. This is done in order to give a chance to change the probabilities of light bulbs that have been in a fixed state for a very long time. When a light bulb stays in a fixed state for a very long time and it turns out that we do not have serious arguments to unfix it, and we have not found the desired situation, it may be time to change something - to unfix this light bulb and see what happens, if the probabilities are given the opportunity to change. But it is worth doing this quite rarely - the constant $ a $ should be made a lot bigger than $ m $, $ m_1 $, $ b $, and $ C $.

After we have unfixed the light bulb, we need to give the probability with it the opportunity to change freely for some time, without giving the algorithm a chance to immediately fix it on the grounds that the probability with it is at a distance less than $ \delta $ from zero or one. For this, immediately after unfixing, we assign the status "unfixable"\ to the light bulb for the time $ C $.

\section {Search for cocyclic polynomials.}

\subsection {What we are looking for.}

In the last section I wrote that we will look for (and give preference to them when adding to $ \Omega $) most discriminated and most semi-discriminated cocyclic polynomials. I think, first, I should explain why this is so, at the same time defining the semi-discriminated polynomials.

First, let's discuss why we are adding new polynomials to $ \Omega $ at all. After all, if we select a fixed set of cocyclic polynomials satisfying the property that if in the situation they all take the correct values, then the situation is consistent and we will be able to obtain a set of probabilities, each of which is equal to 0 or 1, and the situation corresponding to this set is such that all polynomials of our set take the correct values on it, then we have achieved the goal.

Let me remind you that the algorithm for changing the probabilities works in such a way that it maximizes the probability that a random situation from the $ K $ distribution will be consistent. We implicitly estimate this probability based on what values the polynomials take on random situations from $ K $. It is fair to say that we change the values of the probabilities based on an (implicit) estimate of the probability of consistency of a random situation from $ K $. The reason for adding new polynomials to $ \Omega $ is that the more polynomials, the more accurate this estimate is - new polynomials give us additional information.

Let me remind you that at each step the change in each probability consists of several summands - each polynomial brings its own summand - the argument of the given polynomial.

I argue that when adding a polynomial, we should add the polynomial with the expectedly strongest arguments. To explain this, one could refer to the well-known saying that we can learn little from someone whose opinion never contradicts ours (a polynomial with weak arguments says that the current probabilities are good enough, a polynomial with strong arguments says that no, something needs to be changed), but the reader probably wants more.

The reasoning is following. Imagine a situation when we are faced with a choice of which of two polynomials $ f_1 $ and $ f_2 $ to add to the current set of polynomials $ \Omega $. Moreover, the arguments of the $ f_1 $ polynomial are somewhat stronger than the arguments of the $ f_2 $ polynomial. The most optimal would be to change the probabilities using the arguments of all the listed polynomials: polynomials from $ \Omega $, $ f_1 $ and $ f_2 $ (change polynomials based on more information). But since we are allowed to add only one of $ f_i $, it is better to add $ f_1 $, because if we talk about the next few changes, then the result of changes based on $ \Omega $ and $ f_1 $ differs less from the result of the most optimal changes based on $ \Omega $, $ f_1 $ and $ f_2 $, than the result of changes based on $ \Omega $ and $ f_2 $ differ from the result of changes based on $ \Omega $, $ f_1 $ and $ f_2 $, since the first difference is exactly arguments of $ f_2 $, and the second difference is exactly arguments of $ f_1 $.

So, every time you add polynomials with the expectedly strongest arguments. Now I will explain why discriminated polynomials have this property. I will explain this with the example of two linear cocyclic polynomials, one of which, $ f_1 $, is discriminated, and the other, $ f_2 $, is not. $ t (f_1) = t (f_2) = 0 $, among the coefficients of $ f_1 $ there are $ k $ ones, $ 999k $ minus ones, other coefficients are zeros, among the coefficients of $ f_2 $ there are $ 500k $ ones, $ 500k $ minus ones, other coefficients are zeros. All probabilities $ p_i $ are $ 1/2 $.

Suppose that we want to calculate the strength of the argument of $ f_1 $ corresponding to the light bulb $ s $. If the coefficient of $ s $ in $ f_1 $ is zero, then this argument will be zero. If the coefficient is nonzero, then we divide all situations into classes according to the values of the light bulbs with coefficient 1, other than $ s $. Consider an arbitrary of these classes, let $ l $ of light bulbs with coefficient 1, other than $ s $, take the value 1 in this class. The contribution to $ f_1 $, given by $ s $, differs by 1, depending on whether we turn on the $ s $ or not, be the coefficient of $ s $ is one or minus one. This means that the number of bulbs with a coefficient minus one, other than $ s $, which we have the right to turn on, so that $ f_1 $ is greater than or equal to $ t (f_1) $, that is, greater than or equal to zero (we want to evaluate the probability that $ f_1 $ will take a value that is on the side of $ t (f_1) $, on which it will take a value with a lower probability) also differs by 1: say, $ l $ and $ l - 1 $. But the number of subsets of size at most $ l $ in a set of size $ 999k $ is significantly larger than the number of subsets of size at most $ l-1 $ in a set of size $ 999k $. Approximately $ 999k / l $ times as larger. That is, for each of the considered classes, the probability of $ f_1 $ being greater than zero in one of the cases is significantly greater than in the second, and therefore the argument aimed at increasing the probability corresponding to the second case will be quite large.

Carrying out a similar reasoning in the case of the polynomial $ f_2 $, we can also divide all situations into classes according to the criterion of what values the bulbs will take with a coefficient of 1 and see that in most of these equally probable classes, the probability that $ f_2 $ will be greater than zero, for different values of the light bulb $ s $, differ much less significantly, and therefore, in general, the probabilities that $ f_2 $ will be greater than zero for different values of the light bulb $ s $ differ much less significantly than in the case of $ f_1 $ ... Symmetric reasoning suggests that the probabilities that $ f_2 $ will be less than zero for different values of the bulb $ s $ also differ much less significantly than in the case of $ f_1 $ and "less than zero".

By the way, the idea of using the most discriminated cocyclic polynomials correlates with the book ~\cite {Shap}, judging by its title and some of the phrases I read from it. The book consists of various examples of analysis of solving olympiad mathematical problems and, as far as I understand, the main promoted idea is that if we understand where in the problem it is most difficult for us, then this difficulty should be overcame in the first place.

In addition to discriminated cocyclic polynomials, we will be interested in semi-discriminated ones. A semi-discriminated cocyclic polynomial $ f $ is one for which there is a light bulb $ i $ and $ j \in {0,1} $ such that if you replace $ p_i $ with $ j $, then $ f $ becomes discriminated. Then we say that the polynomial $ f $ is semi-discriminated with respect to the light bulb $ i $ and the value $ j $. In other words, it is a polynomial $ f $ for which $ T_j (i, f) $ is close to zero. It is easy to see that the argument (given the initial probabilities) for the $ i $ th light bulb generated by such a polynomial has a good chance of being strong. We will search for semi-discriminated polynomials for each light bulb $ i $ in a way that is no different from the way we search for discriminated polynomials, with the only difference that we will replace $ p_i $ with 0 or 1.

Perhaps, for reasons of saving resources, it would be worthwhile to restrict ourselves to that for each found semi-discriminated polynomial with respect to an arbitrary light bulb $ i $ and the value $ j $, when changing probabilities, using only that of its arguments that corresponds to the light bulb $ i $, and arguments that match other bulbs should not be evaluated or used.

Let us now turn to the question of how one can find the most discriminated (and semi-discriminated) cocyclic polynomials.

I will propose two approaches to finding the most discriminated polynomials. The first approach we will call "the half-space method"\ and the second - "the quadratic function optimization method"\ . When implementing the algorithm, you can combine both of these methods somehow.

Among these two methods, I prefer the quadratic function optimization method.

Both methods will use the selectable parameters, just like in the previous section.

\subsection {Half-space method.}

Note that if we change the value of $ f $ by a constant, then its value in each of the situations and $ t (f) $ will change by this constant. Therefore, the search for discriminated cocyclic polynomials is reduced to the search for discriminated cocyclic polynomials $ f $ such that $ t (f) = 0 $. Note that the cocyclic polynomials $ f $ such that $ t (f) = 0 $ form a vector space. Let's call it $ L $. Note that any situation $ z $ defines a linear function $ H_z $ on $ L $, given by the formula $ H_z (f) = f (z) $. A polynomial $ f \in L $ will be discriminated if, for a random situation $ z $ from the distribution $ K $, either with a probability very close to 1, $ H_z (f)> 0 $, or with a probability very close to 1, $ H_z (f) <0 $.

Let us choose the situations $ z_1, z_2, ..., z_q (n) $ randomly from the distribution $ K $, where $ n $ is the size of the circuit, $ q $ is some polynomial that is selectable parameter. If we find a polynomial $ f \in L $ such that for the vast majority of $ z_i $, $ H_ {z_i} (f)> 0 $ (or $ H_ {z_i} (f) <0 $, for the vast majority of $ z_i $), then $ f $ with a high probability is discriminated.

Note that the set of those $ f $ for which $ H_{z_i} (f) $ is greater than zero (less than zero), forms a half-space in $ L $, for each $ i $. Thus, our task is following: given a polynomial set of open half-spaces, it is required to select a point belonging to as many of these half-spaces, or a point belonging to as many interiors of their complements as possible (I said "interiors"\ , since the boundary hyperplane on which $ H_{z_i} $ is zero should be discarded from complement).

So, how to "pierce"\ as many half-spaces from a given set as possible? Here's the idea. We start from a random point from some natural distribution on $ L $, which is a selectable parameter (there are few requirements for the distribution, we just want the random point not to fall too far from zero too often). We begin to move around $ L $ in jumps. The displacement vector corresponding to each jump is calculated as follows, depending on the current point $ x $ at which we stand. For each half-space $ H $ from our collection, construct a normal $ d_H $ of unit length to the hyperplane separating $ H $ from the complement to $ H $ and directed towards $ H $. So, the displacement vector corresponding to the jump from the point $ x $ is equal to the sum of $ d_H $ over all half-spaces $ H $ from our set that DO NOT CONTAIN the point $ x $, multiplied by some small selectable parameter $ \lambda_1 $ (we multiply by $ \lambda_1 $, again, out of a desire to bring the movement closer to continuous).

I believe that if there are points in $ L $ that pierce a fraction of $ 1- \varepsilon $ of half-spaces of the set, we have a good chance, when moving, to walk through a point piercing, say, a fraction of $ 1-10 \varepsilon $ of half-spaces of the set.

The reason why I think so is following. Let $ x_1 $ be a point piercing a part of $ 1- \varepsilon $ of half-spaces of the collection. Let's color the half-spaces pierced by the point $ x_1 $ in blue, and not pierced - in red. Let $ \delta (x) $ be the fraction of all half-spaces of the set of blue half-spaces that do not contain the point $ x $. Since each such half-space $ H_1 $ contains $ x_1 $ and does not contain $ x $, the normal $ d_{H_1} $ corresponding to this half-space satisfies that its scalar product with the vector connecting $ x $ and $ x_1 $ is greater than zero. So, if the point $ x $, at which we are staying at the moment, satisfies that $ \delta (x) \le 9 \varepsilon $, then we are already at the point piercing the fraction $ 1-10 \varepsilon $ of half-spaces of the set ... If $ \delta (x)> 9 \varepsilon $, then the displacement vector at the moment will consist of more than $ 9 \varepsilon {q (n)} $ normals "directed towards the point $ x_1 $"\ and only $ \varepsilon {q (n)} $ normals corresponding to red half-spaces (you also need to multiply by $ \lambda_1 $). This means that the next move is more likely to bring us closer to $ x_1 $ than to move us away of $ x_1 $. It is unlikely that it will get too long to approach the point $ x_1 $. However, you need to understand that that it's not always that such a movement will lead us to the desired points.

Let's do the following. Let $ m $ and $ k $ be two sufficiently large numbers that are selectable parameters (I intuitively estimate the optimal choice of $ m $ as much larger than the optimal choice of $ k $). The algorithm is following: we select a random point $ k $ times from the distribution on $ L $, which I wrote about above, start from it and make $ m $ steps in the way described above. We will get $ (m + 1) k $ points. We choose the one that is contained in the largest number of half-spaces in the collection. After that we add the corresponding cocyclic polynomial to $ \Omega $.

\subsection {Quadratic function optimization method.}

We start by looking for linear discriminated cocyclic polynomials $ f $ with zero free term, such that $ t (f) = 0 $. That is, in other words, these are simply invariant linear combinations of a certain set of light bulb values, which output 0 in consistent situations. The value that such a linear combination outputs in a random situation is the sum of random variables $ \tau_i $, where $ \tau_i = 0 $ in case when the $ i-th $ bulb is not lit and $ \tau_i $ is equal to the coefficient at the $ i $-th bulb, in case the $ i $ -th bulb is lit.

Recall that the $ \tau_i $ are independent. This means that you can use the Central Limit Theorem with caution. We are interested in the central limit theorem in the Lindeberg form:

\begin {theorem}

Let the independent random variables $ X_1, \ldots, X_n, \ldots $ be defined on the same probability space and have finite mathematical expectations and variances: $ \mathbb {E} [X_i] = \mu_i $, $ \mathrm {D } [X_i] = \sigma ^ 2_i $.

Let $ S_n = \sum \limits_ {i = 1} ^ n X_i $.

Then $ \mathbb {E} [S_n] = m_n = \sum \limits_ {i = 1} ^ n \mu_i, \; \mathrm {D} [S_n] = s_n ^ 2 = \sum \limits_ {i = 1} ^ n \sigma_i ^ 2 $.

And let the `` Lindeberg condition '' be satisfied:
$ \forall \varepsilon> 0, \; \lim \limits_ {n \to \infty} \sum \limits_ {i = 1} ^ n \mathbb {E} \left [\frac {(X_i- \mu_i) ^ 2} {s_n ^ 2} \, \mathbf {1} _ {\{| X_i- \mu_i |> \varepsilon s_n \}} \right] = 0 $,
where $ \mathbf {1} _ {\{| X_i- \mu_i |> \varepsilon s_n \}} $ is an indicator function.

Then
$ \frac {S_n-m_n} {s_n} \to N (0,1) $ in distribution for $ n \to \infty $.

\end {theorem}

Thus, if there are quite a lot of nonzero coefficients and among them there are no those that are much larger than the others, then we can expect that the distribution of the deviation of our random variable from its expectation divided by $ s_n $ (see the theorem) will be close to that indicated in the theorem normal distribution $ N (0,1) $. Thus, if we take the plot of the $ N (0,1) $ distribution, stretch it $ s_n $ times horizontally and shift it by the expectation of our random variable $ \sum {\tau_i} $ (let's call it $ C $), then we we get a distribution that approximates the distribution $ \sum {\tau_i} $.

So, $ \sum {\tau_i} $ has a distribution specified by two parameters: $ C $ (expectation) and $ s_n $ (stretch). If a linear cocyclic polynomial $ f $, with $ t (f) = 0 $ and zero free coefficient, indeed gives a close to normal distribution with parameters $ C $ and $ s_n $, then looking at the plot of the normal distribution, we understand that the extent of its discrimination (of polynomial) is determined by how much less is $ s_n $ than $ C $ in absolute value. $ T (f) $ is determined by the ratio $ s_n / | C | $ and $ T (f) $ is the less, the less this ratio is.

Note that $ {s_n} ^ 2 $ is a positive definite quadratic form on the space of linear cocyclic polynomials (let's call this space $ L $), with a minimum at zero (by definition, $ {s_n} ^ 2 $ is equal to the sum of variances of all $ \tau_i $, but the $ i $-th coefficient is a linear function on $ L $, and the variance of $ \tau_i $, in turn, is equal to the square of this coefficient multiplied by $ p_i (1-p_i) ^ 2 + (1-p_i) p_i ^ 2 $; in this context we consider $ p_i $ as a constant). In turn, $ C $ is a linear function on $ L $ (taking zero value at zero).

The set of points in $ L $ at which $ C $ is equal to -1 forms a hyperplane $ L_1 $ (unless, of course, $ C $ is not identical zero).

The elements of the space $ L $ that we add to $ \Omega $ are following.

First of all, this is the point $ h $ in $ L_1 $, at which the quadratic function $ {s_n} ^ 2 $ reaches its minimum (on $ L_1 $). This is the very point that is the estimate for the most discriminated cocyclic polynomial in $ L $.

Intuition tells me that along with this polynomial, in $ \Omega $ it is worth adding a few more slightly less discriminated polynomials. I propose to add polynomials corresponding to the following points. Let us write the quadratic function corresponding to $ s_n $ as a function on $ L_1 $, in some orthonormal coordinate system on $ L_1 $, centered at the point $ h $. The result is a quadratic function with a zero linear part and a nonzero constant part. Such a quadratic function can be reduced to the principal axes.

Consider the set of lines on $ L_1 $ passing through $ h $ corresponding to the given principal axes. The points we need are the union of pairs of points along these lines, in which our quadratic function is equal to $ \theta $, where $ \theta > 0 $ is a selectable numerical parameter, small in absolute value. You shouldn't look for internal logic in this particular choice of points other than $ h $, I just want to add a set of points surrounding the point, which is the estimated most discriminated cocyclic polynomial, at all sides. I decided that the indicated points at the main axes would be the most natural choice ...

The search for the minimum point of a positive quadratic function and the presentation of the quadratic form to the principal axes is carried out by standard algorithms in polynomial time.

What can be said about cocyclic polynomials of greater degree, with possibly nonzero free coefficient?

Let $ \tau_i $ be a random variable corresponding to the $ i $-th monome (equal to zero if the light bulbs corresponding to the given monome are not all lit and equal to the coefficient of this monome if all of them are lit). The problem is that the $ \tau_i $ are not independent this time, so the Central Limit Theorem can not be applied.

But the Internet is full of generalizations of the central limit theorem to cases when the independence condition is somehow weakened, and our random variables $ \{\tau_i \} $ are close to independent, since two monomials of a cocyclic polynomial of small degree taken at random with high probability do not intersect in light bulbs, which means that the random variables corresponding to these monomials are independent with high probability. Therefore, it is possible that the distribution of the random variable $ \sum {\tau_i} $ interesting to us is still very often close to normal and we can apply the same method.

But I like the other approach more. We considered light bulbs, each of which either lights up or does not light up and each monomial is calculated as the product of zeros and ones corresponding to the corresponding light bulbs, multiplied by some factor. You can do a little differently by assigning one large light bulb to each monome. This light bulb will no longer depend on the adjacent gates, it will depend on the union of the gates of all the bulbs corresponding to a given monome, and if we want to understand which bulbs light up at a selected set of gate values, then a specific bulb will light up exactly when each of bulbs corresponding to this monome lights up.

And the concept of a situation can be given meaning as a matching of zeros and ones to each of the bulbs of this considered set of bulbs, each of which is determined by a set of no more than $ d $ sets of adjacent gates and the values chosen for these gates.

The case when the number of such sets is equal to zero corresponds to the free term of polynomial.

That is, each lamp polynomial in the old sense corresponds to a linear lamp polynomial on the considered set of bulbs, the value of which, calculated for any set of gate values of the circuit, will be the same.

Since the values for the same sets of gate values coincide, the invariance of the lamp polynomial on the considered set of bulbs can be checked by checking the cocyclicity of the corresponding lamp polynomial in the old sense.

The algorithm in the case of linear polynomials on the extended set of light bulbs looks exactly the same, but the random variables $ \tau_i $ corresponding to the bulbs of the extended set will be independent, which opens the way for applying the Central Limit Theorem and searching for the expectedly most discriminated cocyclic polynomials the way discribed above.

Sometimes it is very useful to add to $ \Omega $ cocyclic polynomials in which there are few nonzero coefficients for monomials (or bulbs of an extended set of bulbs) with the probability of becoming one separated from 0 and 1 (in particular, polynomials in which there are few nonzero coefficients for monomials / bulbs of an expanded set of bulbs with a probability of becoming one other than 0 and 1). But such polynomials are no longer discussed in the context of Central Limit Theorem and it cannot be argued that the corresponding distribution is close to normal (there are terms that make a significant contribution), therefore, we shouldn't wait for our algorithm to find such a polynomial. To partially compensate for this, one can choose the constant $ Q $, iterate over all subsets of the set of monomials/bulbs of the extended set of bulbs with a probability of becoming one, different from 0 and 1, of size at most $ Q $, and for each such subset find cocyclic polynomials (and if the find will suit us - add it to $ \Omega $) in the constant-dimensional space of cocyclic polynomials whose coefficients are zero outside this subset. You can search for them, for example, using the half-space method.

\section {On the estimated complexity of the algorithm.}

Since the algorithm is stochastic, one should not expect explicit upper bounds on the running time for it. We can only talk about estimates of the average work time, for some variation in the definition of the average complexity. It is assumed that the ideal algorithm I am striving for will work on average in a time equal to the polynomial of the length of the input string and the running time of the Turing machine (circuit height) multiplied by the exponent of the record length $ m $ of the Turing machine. The exponential dependence on the TM size appears at least due to the fact that in each local neighborhood of the gate we use an exponential in $ m $ number of bulbs. The multiplicative constant in the exponent of $ m $ almost certainly allows for the possibility of its rather serious decrease.

Moreover, in cases that come from practice, the dependence can often be made polynomial in $ m $. This happens because often for TMs that come from practice, the calculation on this TM can be represented as a circuit, each gate of which is calculated from no more than 8 other gates (the constant 8 was chosen very conditionally). Having presented the circuit in this form, one can afford to consider and use only light bulbs that depend, say, on no more than 12 gates of the circuit (for each light bulb, these gates can either be located next to each other or be scattered over different parts of the circuit). But, we must understand that winning in resources, limiting ourselves in the size of the bulbs used, we lose something. In particular, the strength of our evidence system is diminishing somewhat.

\section {Big string writing problem.}

Now let's talk about the big string writing problem: it is required to construct an algorithm $ S $, according to the circuit $ M $, constructing an algorithm $ L $, which, from a string $ A $ of fixed (depending on $ M $) length, constructs a string $ B = L (A ) $ of fixed (depending on $ M $) length such that $ M (A, B) = 1 $.

Consider a Turing machine representation of $ L $. We will restrict ourselves a little, deciding to look only for such Turing machines $ L $ that work for the time bounded from above by the polynomial $ p_1 $ in terms of the sum of the total lengths of $ A $ and $ B $ and the size of the circuit, the head of which does not move away from the initial position during operation by a distance exceeding the polynomial $ p_2 $ in terms of the sum of the total lengths of $ A $ and $ B $ and the size of the circuit, and such that the alphabet that this machine operates with and the number of its states are limited by the constant $ C_3 $. We do not limit ourselves too much, since we can apply our algorithm many times for increasingly large polynomials
$ p_1 $ and $ p_2 $ and $ C_3 $ constant. This means that $ L $ can be represented as a polynomial-sized circuit.

Let me remind you how we transform the Turing machine into a circuit. Let's mark each cell of a rectangle of the corresponding size with coordinates $ (x, y) $ with the state of the tape in the tape cell with the number $ x $ at the time $ y $, if we run our TM on some string. The state of the tape consists of the symbol written in this cell at the moment, a bit that means the presence of the TM head in this cell at the moment, as well as the TM state encoded with zeros and ones at a given moment. Note that the labeling of a cell with coordinates $ (x, y) $ is determined by the labels in cells with coordinates $ (x-1, y-1) $, $ (x, y-1) $, $ (x + 1, y- 1) $. Thus, we can compose a Boolean circuit, each gate of which corresponds to a bit of a mark of a certain rectangle cell, and the value of this gate is calculated in a natural way from all the gates of the marks of three rectangle cells located on a row one below directly under this cell, slightly to the left or slightly to the right.

Thus, to find $ L $, it is enough to find the circuit, $ L $ will be determined unambiguously after that. Since the state of the tape at a given point and time is encoded by a constant number of bits - the gates of the circuit, all the gates of the circuit are divided into a constant number of types: each type consists of a set of gates corresponding to each other and has one gate in each cell. And gates of the same type are calculated from the gates of the cells located under them in the same way. Thus, it is enough for us to construct a function for each of the types, via which the value of each of the gates of this type is calculated from the values of the gates of the cells below it.

And each of these functions is naturally determined by the set of zeros and ones in a quantity equal to the number of possible values that the bits of three consecutive cells can take.

Thus, $ L $ is defined by the string $ N $, consisting of such sets of zeros and ones, for all types of gates in the circuit. If such a string does not correspond to any Turing machine, its consideration is still pertinent: the Turing machine cannot be determined according to the corresponding circuit (it of course can, but in another, slightly less trivial, way), but this circuit can as well be used as an algorithm for finding the string $ B $ by string $ A $.

Moreover, you can give yourself more freedom and allow functions corresponding to different gates of the same type to be different (and then the length of the string defining the algorithm $ L $ - we will allow the freedom of speech and also call it $ N $ - will be multiplied by the size of a circuit). In this case, perhaps we will get a circuit of big complexity.

One way or another, the algorithm $ L $ can be represented as a string $ N $ (this was already clear, I just wanted to emphasize that this can be done in different ways and since these ways are quite different in nature, the efficiency of the algorithm $ S $ may depend on the chosen way) and we will look for the algorithm $ L $ in the form of the string $ N $.

Let us denote by $ N (A) $ the string $ B $, which is built from the string $ A $ by the algorithm that is encoded by the string $ N $. Let $ M_1 $ be a naturally constructed Turing machine that takes two rows $ A $ and $ N $ and satisfies the condition $ M_1 (A, N) = M (A, N (A)) $.

Thus, we have brought the task to following form. Given a machine $ M_1 $, you need to construct a string $ N $ of fixed length, such that for any string $ A $ of a fixed length, $ M_1 (A, N) = 1 $.

My approach is following. We will maintain two lists: the first consists of algorithms $ \{A_i \} $ that construct the first argument of $ M_1 $ from the second argument, the second consists of candidates $ \{N_i \} $ for the second argument of $ M_1 $. We will find and add items to each of these lists.

We initialize the lists as you like. Each time, adding an element to the second list, we select some sufficiently large, polynomial in size, subset $ A_ {i_1}, A_ {i_2}, ..., A_ {i_k} $ of the first list and look for a string $ N $ such that $ M_1 (A_ {i_j} (N), N) = 1 $, for any $ j $.

It is easy to see that such a problem can be formulated as a small string writing problem. Let's try to solve this problem. If it turns out that such a string $ N $ does not exist, then we can conclude that we will not be able to find a solution to this problem: for any string $ N $, there is a string $ A $ for which $ M_1 (A, N) = 0 $. If the required string $ N $ is found, then we add it to the second list. If we cannot solve a given instance of a small string writing problem, then we do nothing and move on.

Each time, adding an element to the first list, we select some sufficiently large polynomial in size subset $ N_ {i_1}, N_ {i_2}, ..., N_ {i_k} $ of the second list and look for an algorithm $ A '$, such that $ M_1 (A '(N_ {i_j}), N_ {i_j}) = 0 $, for any $ j $.

It is easy to see that such a problem can be formulated as a small string writing problem. (If, again, we restrict ourselves and represent $ A '$ as a string of fixed polynomial length $ l $.) Let's try to solve it. If such an algorithm $ A '$ can be found, we add it to the first list. If not, do nothing and move on. It should be noted that here it is necessary not to overdo with the value of the parameter $ l $: if we make $ l $ large compared to the sample size $ k $, then we can get an algorithm that simply iterates over all $ N_ {i_j} $, choosing one of them , which coincides with the input $ N $ and outputs the string $ A $ stored in advance for the given instance, such that $ M_1 (A, N) = 0 $. Therefore, some kind of regularization is needed, for example, in the form of prohibiting $ l $ from being large.

We will add items in the specified way to the first and second lists in turn. Our goal is either someday to find the required string $ N $ in the second list, for which, for any string $ A $, $ M_1 (A, N) = 1 $, or to find the algorithm $ A '$ in the first list, for which for any string $ N $, $ M_1 (A '(N), N) = 0 $.

We will say that the string $ N $ wins the algorithm $ A '$ if $ M_1 (A' (N), N) = 1 $. In case if $ M_1 (A '(N), N) = 0 $, we say that $ A' $ wins $ N $.

The logic of the algorithm is to get algorithms in the first list that defeat as many strings of the second list as possible, and in the second list - strings that defeat as many algorithms of the first list as possible. So we hope to find an algorithm that conquers all possible strings, or a string that conquers all possible algorithms (and therefore is the desired one).

Certainly, since we want to find such a string or an algorithm, we should from time to time go through all the elements of both lists and check for each of them whether it is the desired one (obviously using the algorithm for a small string writing problem).

I have omitted until now how exactly we should collect samples from $ k $ elements of one list or another. We will choose these $ k $ elements randomly, but different elements will have different probabilities $ p $. And the probability of a given element will monotonically depend on the strength $ r $ of this element. (I do not specify exactly how it will depend, this function is a selectable parameter.) The strength of an element is defined as the probability that an element will win an element of another list, where the probability $ q $ of an element of another list will be chosen taking into account its strength. Now I will explain how to give a logical meaning to this definition.

The set of values of the strength $ r $ and the probability $ q $ on all elements of both lists is a system that is self-balancing in time.

In addition to the probability $ p $ and the strength $ r $ of the elements of both lists, we will support such parameters as the ideal of strength $r'$, the distribution $ Q $, consisting of the probabilities $ q $ of choosing an element of each list in the distribution $ Q $, and the ideal $ q'$ of the probabilities of choosing an element in the distribution $ Q $.

The strength of $ r $ is a time-varying quantity, parallel to the main part of the algorithm, which is adding and removing items from lists. The ideal of strength $ r'$ of an element is equal to the probability that an element of the opposite list loses to this element if this element of the opposite list is selected from the distribution $ Q $, where the probability $ q $ of choosing each element is a time-varying quantity, like $ r $. The ideal $ q '$ of the probability of choosing an element in the distribution $ Q $ is a monotonic function of the strength $ r $ of this element, which is a chosen parameter.

In one step of the parallel to the main part algorithm that changes the values of $ r $ and $ q $, we change each value $ q $ by $ (1 - \varepsilon) q + {\varepsilon} q '$, and each value $ r $ by $ (1 - \varepsilon) r + {\varepsilon} r '$, where $ \varepsilon $ is a small in absolute value selectable parameter.

As for the main part of the algorithm, I described, how we add elements to the lists, it remains to add that from time to time, in order to maintain the optimal size of the lists, we throw out the elements of the least strength $ r $.

In this algorithm, minor details of which have been omitted, echoes the already mentioned idea from ~\cite {Shap}.

Perhaps it was worthwhile to limit ourselves to always choosing an algorithm from the first list that always produces the same fixed string, in other words, arrange everything so that the first list contains the strings themselves, and not the algorithms looking for these strings (after all the second argument of $ M_1 $ in our case always encodes some algorithm, and looking for an algorithm that builds a string by the algorithm is not very natural). But perhaps, if we leave everything as it is, the algorithm will often be more efficient (we can say with a fairly high degree of certainty that it would be more efficient if we did not have the assumption that the second argument of $ M_1 $ - the string $ N $ - comes from some algorithm, and the machine $ M_1 $ is a machine of this rather special kind).

\section {Further work.}

In this section, I will often talk informally. I will give a number of ideas that I plan to work on further and a rather crude scheme for their implementation (namely the scheme, not the implementation itself). Perhaps this section is worth reading if the reader plans to join me, or, on the contrary, wants to know what it is about, and perhaps not to return to this in the future. Otherwise, it may be worth waiting for further works on this topic, in which, I hope, the topic will be covered in more detail.

\subsection {Feature space.}

In the previous sections, we considered all possible arrangements of zeros and ones in the gates of the circuit and calculated for each such arrangement the values of the light bulbs: each such value was calculated as a function of the values of the constant number of sets of adjacent gates (we had an indicator function for a specific set of values, but you can do not limit yourself so and use any propositional formula). But what if go further and start wondering about the values of the functions of gate assignments that are computed not locally?

      If we are talking about circuits that came from a Turing machine, then the set of gate values for such a circuit can be viewed as a kind of two-dimensional string. And on this two-dimensional string, a Turing machine can work, which is specially designed to work with two-dimensional strings: each time the head of such a machine is shifted left, right, forward or backward, its new state and displacement are determined by the old state and the characters written in the cells, each coordinate of which differs from the corresponding coordinate of the head by no more than one.

      Thus, each such "two-dimensional" Turing machine corresponds to a function of assigning values to gates - the answer given by such a machine. Moreover, we can represent such a Turing machine in the form of a three-dimensional circuit, it will have its own gates and its own light bulbs. And we can consider the cocyclic polynomials for this three-dimensional circuit in the same way, changing the probabilities with its light bulbs (although this time, we do not have a fixed value that such a circuit should give as an answer) and this can help to restore the values of the bulbs of the original circuit (since they are also light bulbs of the new circuit, which means they can participate in the cocyclic polynomials of the new circuit). I wrote about this when I was talking about a cage with an additional construction.

      On this three-dimensional circuit, we can compute something on a "three-dimensional" Turing machine, and so on.

      Each light bulb and gate of any circuit constructed in this way corresponds to the function of the input string: from the input string, you can calculate the history of calculating of output of the original circuit (the values of all gates), from the history of computing of output of the original circuit, you can calculate the values of all the bulbs on the original circuit. We can also calculate the history of computing of output of any receiving it as an input three-dimensional circuit. This history, in turn, can be used to calculate the values of all corresponding to it bulbs, and so on.

      Note that each such function of the input string can be written as some Turing machine applied to the input string and a certain number of natural numbers written in the unary number system.

      For example, the light bulb of the original circuit, depending on the neighborhoods of two cells of the original circuit, is encoded by four numbers - the coordinates of these two cells in the circuit rectangle - and by a Turing machine, which first calculates the entire history of work of the original Turing machine (implemented by the original circuit) from the input string, then via given four numbers finds all the values of the gates from the neighborhoods of the two required cells, and from them finds the required value, applying the required propositional formula.

      Let's call a particular feature a pair consisting of a Turing machine and a set of certain natural numbers. We have just realized that any function of interest to us is given by a particular feature.

      Let us call a generall feature a pair consisting of a Turing machine and one (usually small) natural number - feature arity. A generall feature specifies not one function, but a whole class of functions of the original string, parameterized by a set of sets of natural numbers in an amount equal to the arity of a given generall feature: indeed, to each such set of natural numbers there corresponds a particular feature, which corresponds to the same TM and exactly this set of natural numbers. We will call such a particular feature the specification of this general feature.

      Thus, the small string writing problem can be formulated as follows. Find a string whose given feature is equal to one. (This feature can be considered both particular and general arity zero.)

\subsection {Navigation in the feature space.}

At any moment in time, the algorithm will maintain a set of particular features, which we will call consciousness. (I call it consciousness because there is an analogy with human consciousness. These are the particular signs "of which the algorithm is thinking at the moment.")

      When our algorithm works, it will add new particular features to consciousness and throw out some particular features from it. At the same time, for each of the particular features of consciousness (and for some other particular features), the probability (the probability of holding of the feature) will be maintained, which, as the reader has probably already guessed, we will change over time.

      At any moment, the attention of the algorithm will be focused on a set of particular features, which we will call the locus of attention.

      Each time we select particular features in the set, which we want to make the locus of attention (on which we want to focus attention) to a large extent randomly, but at the same time, we are interested in the that there are quite a few particular features from consciousness in the locus of attention, as well as in that there are many connections between the particular features of the locus of attention. By connections we mean simply clauses - propositional constraints on the set of particular features of constant size. We want all this in order to be able to determine some values of particular features of the locus of attention from others. We will determine it in a probabilistic sense - we will bring the probabilities of some features closer to zero or one.

      Now let's talk about how exactly we will change the probabilities. We will change the probabilities of those features that are in the locus of attention. We will change them in the same way - with the help of cocyclic polynomials. It is only necessary to generalize the concept of a cocyclic polynomial to the case of a feature space. For each such polynomial there is a set of so-called basic features, as well as a set of so-called lamp features. Each lamp feature is calculated for no more than $ d $ groups of adjacent basic features ($ d $ is a constant) using a propositional formula. A protected lamp feature is one that cannot light up (to light up - this means to take the value 1) if the basic features take such values that they should take on some input string. We consider all possible arrangements of zeros and ones in basic features, just as we considered all possible arrangements of zeros and ones in gates in the previous sections. A cocyclic polynomial is an arrangement of rational numbers on lamp features, such that the sum of the numbers corresponding to the lamp features that are lit up does not depend on what values the basic features take. We would also like to check the cocyclicity of a polynomial locally. Of course, we also need to define what "adjacent features" mean and how exactly to check cocyclicity locally (most likely, in the same way, but questions may arise if in a small neighborhood of a feature there are more other features than we would like) , this remains to be done.

      We will change the probabilities of unprotected lamp features in the same way as in the previous sections. The probabilities of protected lamp features are always zero. For the current locus of attention, we will try to identify as many sets of basic features as possible, a set of lamp features corresponding to which intersects as strongly as possible with the locus of attention. One of the options for doing this is to declare a subset of the locus of attention as basic features and to include these basic features in the set of lamp features. But perhaps, in some cases, you can try to select the necessary basic features outside the locus of attention. In general, it is likely right to choose the locus of attention every time so that it includes many constant sets of features, such that many of the propositional formulas calculated from them also belong to the locus of attention. And we combine the sets of unprotected lamp features for all the found sets of basic features, add them to the locus of attention, if they are not there yet, and change the probabilities for them in the same way as we did in the previous sections: looking for the most discriminated polynomials, a set of basic features of which is one of the found sets, we add them to the set of processed polynomials $ \Omega $, counting the total arguments for each of the lamp features of the set we are considering, changing the probabilities according to these total arguments, fixing and unfixing values when needed.

      I wrote that we strive to ensure that there are many connections between the particular features of the locus of attention. Why do we need many connections? If there are few connections (few clauses) on the set of particular features of the locus of attention, then, in this case, the cocyclic polynomials are not informative: a lot of lamp features will not be protected (if they hold - take the value one - it will not imply a contradiction), which means, say, if we we want to get a cocyclic polynomial that always gives one, it will be very difficult for us to "hide" most of its positive coefficients in the set of protected monomials. And in general, from general considerations, it is clear that if there are few connections, then it will be difficult to restore some values of particular features from others.

      Thus, at any moment of time we keep in our memory the probabilities for all the particular features of consciousness, as well as for all the features that have visited the locus of attention at least once and were not thrown out of consciousness.
      
\par \bigskip
\par \bigskip
\par \bigskip

What particular features are we going to add to consciousness? We will be interested in features that have two properties to a large extent. Adding a feature, holding, to a large extent, only one of these properties (and the second - to a mediocre degree) is also possible, but the possession of the second property is encouraged. We are interested in some summary indicator.

      We add to consciousness, first, the most definite signs. A feature is considered definite if its probability is close to zero or one. A feature can become definite if, for example, we focused on a certain set of particular features containing this feature, and its meaning was determined unambiguously.

      Another case, which is just as interesting for us, is when a feature has visited the locus of attention several times and each time we found weak arguments to make it, say, a one. But, if you like, the total argument for all these loci of attention to make it a one turned out to be quite large. Therefore, we brought the probability of its holding closer to one.

      When the arguments to make a feature zero or one are consistent, this is a reason to pay attention to the feature. In general, it seems to me that the meaning that the word "consistency" (consensus, conformity, alignment, concurrence - I don't know which of these words fits more to the russian word "soglasovannost'") is fraught with is very important for artificial intelligence. We can also concentrate attention on the set of particular features that lie entirely in consciousness. And if, while focusing on such a set, the arguments of the corresponding cocyclic polynomials say that it is necessary to increase the already large probabilities and decrease the already small ones, then this is always nice (and if not, then something needs to be changed). We are pleased when different loci of attention "speak about the same thing," that is, they are consistent with each other.

      The second property that is welcome in order to add a particular feature to consciousness is the influence of the trait. So far I understand less about the nature of this property than about definiteness and so far I am not able to strictly define it. A feature is considered influential if it has many connections with features of consciousness. In other words, if by its value it is possible to determine the value of many features of consciousness. Features that can be calculated from the features of consciousness in many natural ways often become influential. Features that participate in many clauses (propositional restrictions) with features of consciousness also often become influential.

      There is a symmetry between definitness and influence: definite features are those that can be reconstructed from the features of consciousness, and influential are those from which features of consciousness can be reconstructed.

      Particular features, in the values of which we lose interest, we throw out of consciousness. (Again, it remains to clarify what these features are.)
      In general, I believe, that it is necessary to learn to select from the set of particular features the smallest possible subsets of the most influential features, by which the rest of the particular features of the set are reconstructed. If we select such a subset, it will be possible to remove from consciousness the rest of the features of the set as unnecessary, freeing up memory for new features.

      I would like to point out the not yet fully thought-out possibility of adding features to consciousness, for which we do not indicate at all a way for calculating them from the input string and parameters, as though leaving the finding of this way for later. Initially, we only possibly indicate the connections in which such a feature is with other features of consciousness.

\par \bigskip
\par \bigskip
\par \bigskip

It would be nice to somehow outline the set of particular features, from which we collect features into the locus of attention. If we want to briefly and roughly describe the set of particular features that we consider candidates for inclusion in consciousness, then these are those particular features that are briefly calculated based on particular features of consciousness.

      Let's try to say it less roughly. We will be interested in "refined general features". A refined general feature (hereinafter referred to as RGF) is a general feature, however, the set of numbers supplied to the input of its Turing machine is limited by another, subordinate Turing machine. In other words, we are interested in the value of the first TM, by far, not on all possible sets of numbers, but only on those that satisfy a certain property given by the subordinate TM.

      In addition to consciousness itself, we will maintain "set of central RGF" and "set of peripheral RGF". Set of central RGF and set of peripheral RGF are the sets og RGF that will change over time. The particular features covered by the union of central RGF and peripheral RGF are precisely the set of particular features from which we collect the features in the locus of attention.

      The set of central RGF satisfies that the set of particular features covered by the central RGF substantially intersects with consciousness. In some sense, it can be considered a prototype of consciousness. When we see that one of the peripheral RGF began to overlap strongly with consciousness, we add it to the set of central RGF.

      The set of peripheral RGF can be characterized as a set of features that are shortly computed from the central RGF. The property of being shortly computable from central RGF can be defined formally as follows. Since the peripheral RGF is a feature, it calculates something from the input string and a set of numbers in the unary number system. The peripheral feature operates as follows. It takes several central RGFs $ U_1, U_2, ..., U_k $ and calculates all values of each of them on the input string and all possible sets of numbers satisfying the property specified by the corresponding subordinate TM. I will make a reservation that it will be problematic to calculate something on absolutely all sets of numbers satisfying this property, since there can be infinitely many of them, so we will restrict each of these numbers from above by polynom of the numbers fed to the input of this peripheral RGF. Of all the calculated values of the central RGF, TM, corresponding to our peripheral RGF makes a string in a natural way: the set of considered inputs for each of the considered central RGF forms a rectangular parallelepiped of dimension equal to the arity of the corresponding central RGF, we number its nodes in a natural way and put in turn in a string the values of the corresponding central RGF found for all these nodes, or special symbols if the corresponding cells of the parallelepiped do not satisfy the property specified by the corresponding subordinate TM. There will be several strings, one for each taken central RGF, and we concatenate them, inserting separators between them. So, according to the string obtained in this way and natural numbers - the inputs of the peripheral feature, the output of the peripheral feature is calculated using a small Turing machine $ R $ (that is, written using a small number of bits). The feature obtained in this way will be denoted by $ (U_1, U_2, ..., U_k) \cdot R $.

      Thus, to set the peripheral RGF, you need to specify a set of central RGF and TM of small size. It is likely that among small TMs one TM can be preferred over others, and among the peripheral features, only the most preferable, most meaningful, and most relevant ones can be left, and the rest can simply be not considered. If not, all small TMs will have to be iterated through.

      We see that the set of peripheral RGF is determined by the set of central RGF. Therefore, when we change the set of central RGF, the set of peripheral ones also change. As I said, the RGF passes from the set of peripheral to the set of central ones if many of the particular features it covers pass into consciousness. If there are few particular features of consciousness among the particular features covered by some central RGF, we throw this RGF out of the set of central ones.

      Another thing that should be negotiated is that all particular features corresponding to the gates of the original circuit always remain in consciousness and we never throw them out of there. I hope, I sucseeded in conveying the intuition behind this all. Consciousness is such a cloud that moves through the space of features, changes their probabilities and always contains the gates of the original circuit (the gate, which is considered the output of the circuit, also always remains in the cloud, but the probability with it is always equal to 1 - it is prohibited to touch it). When the probabilities at the gates of the original circuit are all equal to zero or one, and no clause of the original circuit is violated, the work of algorithm ends.

\subsection {Gluing.}

Here I will talk about an idea related to the fact that there are often very many ways to calculate the same particular feature.

      I will reason in the context of the big string writing problem. More precisely, the problem to which we have reduced the big string writing problem in the section on the big string writing problem: given a TM $ M $ that takes two strings $ A $ and $ B $ of fixed length as input and it is required to find such a string $ B $, so that $ M (A, B) $ is equal to one for any $ A $. (When reduced, the string $ B $ encoded some algorithm, and the machine $ M $ was of a rather special kind, but the described problem is of independent interest in the general case.)

      In the context of this task, it is necessary to slightly modify the concept of a feature. This time, in addition to the input string $ B $ and several natural numbers - parameters, the Turing machine calculating the particular feature will take the string $ A $ (the first argument of the machine $ M $) as a parameter. A general feature is just a Turing machine that calculates the value of a feature, without specifying numbers - parameters and the string $ A $. Accordingly, a refined general feature is a general feature for which the set of parameters on which it is defined is limited by some subordinate TM.

      The algorithm works like this. We collect a large enough polynomial sample of strings $ \{A_i \} $, which $ M $ takes as the first input. And for each of these strings in the way described above, we solve the problem of finding a string $ B $ - a small string writing problem, but with one nuance: for different $ \{A_i \} $ we identify the features corresponding to the bits of the input string $ B $. That is, we obtain, say so, many floors, each $ \{A_i \} $ corresponds to its own floor, on each of the floors its own consciousness travels; the features corresponding to the bits of the input string $ B $ play the role of heating pipes, each of which runs through each of the floors. When the probabilities change, everything happens in exactly the same way, except that the probabilities at the features corresponding to the bits of the input string $ B $ take arguments from all floors at once and the probability at each such feature is always the same for all floors. In the rest, the change in probabilities at particular features on each floor happens autonomously.

      The idea is that if at some point we find that the probabilities for the corresponding particular features covered by two general refined features $ U $ and $ V $ are correlated, this means that it is possible that pairs of the corresponding particular features covered by these RGF can be identified ... By corresponding we mean particular features, representing the application, respectively, TMs of $ U $ and $ V $ to the same set of parameters (that is, to the same string $ A $ and the same numbers - parameters) ...

      In more detail, instead of storing the union of consciousnesses across all floors (hereinafter I will call this union simply consciousness) as a set of triples of the form (TM, string - the first input of $ M $, a set of numeric parameters), we will store an abstract set $ N $, for each of the elements of which one or more particular features are specified, the values of which are the same. And for each element of $ N $ is assigned the probability (the same) that the corresponding features will turn to one. In turn, we will not store the set of central and peripheral RGFs either, instead we will store an abstract set $ N '$, for each of the elements of which one or more refined general features are indicated, the values of which coincide on all possible values of the parameters.

      There is an expanded version of consciousness (which we do not store) and there is a set $ N $. Each particular feature of consciousness is projected into one of the elements of $ N $, however, for each element of $ N $, we store in the attached list of elements projected into it not all the elements projected into it, but, perhaps, only a small part of them.

      Thus, on the glued version of consciousness - the set $ N $ - we have new cocyclic polynomials that did not exist before and from which a lot of useful information can be extracted - we are not considering basic and lamp features, but the basic and lamp elements of $ N $.

      Similarly, there is an expanded version of the union of central and peripheral RGF (which we do not store) and there is a set of $ N '$. Each particular feature of the expanded version of the union of the central and peripheral RGF is projected into one of the elements of $ N '$, however, for each element of $ N '$ , we store in the attached list of elements projected into it, not all elements projected into it, but, possibly, only a small part of them.

      If a particular feature is an RGF specification, then we say that the $ N $ element into which the particular feature is projected is the specification of the $ N '$ element into which this RGF is projected.

      For each pair (or for some set of pairs) of $ N '$ elements, we keep track of a certain number of corresponding pairs of $ N $ elements, which are specifications of these $ N' $ elements. And if the probabilities for the corresponding specifications are sufficiently strongly correlated (in the sense that, roughly speaking, most often either they are both close to one, or both are close to zero), we decide to perform a large-scale gluing operation.

      Let's talk about this operation in more detail. This operation consists, firstly, in glueing these $ N '$ elements and glueing the corresponding pairs of their specifications in $ N $. But it was possible to stop at this, why am I talking about this operation as something on a larger scale? Because from the equality of two features, the equality of very many other pairs of features usually follows semantically. For example, we decided to glue the features $ U $ and $ V $ and all their corresponding specifications and there are two another features $ T_1 $ and $ T_2 $. This means that it would be correct to glue, in particular, in the same way, all pairs of features (and, of course, pairs of corresponding specifications), which are, in one case, a propositional formula in terms of $ U $, $ T_1 $ and $ T_2 $, and in the other, the same propositional formula in terms of $ V $, $ T_1 $ and $ T_2 $. Next, you can produce gluing of pairs of features, the equality of which follows from these identifications, and so on. One gluing can be followed like an avalanche by many other gluings and a kind of collapse of the space $ N $ into a smaller space (as well as the space $ N '$) will take place.

      We could make these identifications (gluings), following like an avalanche after one "trigger" gluing, in the same way: after trigger gluing, the considered pairs of features that are destined to be glued are expected to correlate, which means they can be glued in a general order. But it seems to me that these subsequent glues can be done more efficiently. For example, a propositional formula in terms of $ U $, $ T_1 $ and $ T_2 $ and the same propositional formula from $ V $, $ T_1 $ and $ T_2 $, which was discussed above, can be glued together immediately, without waiting for the corresponding probabilities to start correlating.

      This is the direction of further work, which I currently consider the main one - how to correctly make large-scale gluing of elements of $ N $ (and elements of $ N '$). It is unlikely, of course, that the gluings will be grouped into clearly distinguished groups, each of which will consist of trigger gluing and many other gluings arising from it; most likely, such groups will mix in a single stochastic process.

      It is worth clarifying that there are three different situations in which we should glue features (between which, in our context, it might be pointless to draw a border). First, two particular features can, in principle, always always coincide, regardless of the state of some probabilities that we support. Secondly, the coincidence of particular features may follow from the fact that the probabilities for some of the features turn to one or zero, or from the fact that some features have already been glued together. And thirdly, the coincidence of features may not follow neither from the structure of mathematics, nor from the properties of the system of probabilities, nor from something else, but we still decide to identify these features, since they, for a given probability distribution on features, coincide with a large probability or even rather weakly correlated; the reasons to force the equality of such features by gluing them together are following.

      We glue the features together to reduce the number of elements in the set $ N $, so that many other features can be added to the set $ N $ - our memory is limited.

      When we glue together features that are only fairly weakly correlated, we certainly lose some freedom, but it pays off with the opportunity to add many new features to $ N $ and start processing them.

      Perhaps the gluing needs to be organized more gently. Let me explain. In the described approach, we sharply glue features together as soon as their correlation exceeds some lower bound. But it is possible, that this lower bound can be raised and the "gluing pressure" can be used: when the probabilities change, the arguments described in the previous sections to change the probabilities in one direction or another will be proposed not only by cocyclic polynomials, but also by pairs of refined general features. Each pair of refined general features will impose their arguments on each of the specifications of these two features, aimed at ensuring that the specifications of these two features with the same parameters take the same value as often as possible, that is, that the probabilities for the specifications of these two features correlated as strongly as possible (each argument moves the probability with its own feature towards the probability with the corresponding feature). The peculiarity is that when features are weakly correlated, the force to make them correlate as strongly as possible is weak (that is, the corresponding arguments are weak). When traits are strongly correlated, this power is strong. This force, forcing to correlate features, is the stronger, the more these features are already correlated. When features begin to correlate strongly enough, we glue them together (the lower correlation limit at which we glue features should be raised compared to the lower correlation limit at which we glue features used in the approach without gluing pressure). When using the gluing pressure, the gluing of the features is smoother - the leaps generated by the gluing change the situation less strongly.

      There is some reason why, perhaps, we should not introduce an abstract space $ N '$, but work with refined general features as they are. The reason is that two RGF often correlate not on their entire domain (determined by the subordinate TM), but on a certain subset of their domains, set by some Turing machine. Anyway, the domains themselves may be different. This means that we cannot just take and glue these two RGF. And we would like to glue their specifications on the indicated subset of their domains (if this subset is quite simple). This means that it may be worth keeping the RGF in its original form, without starting the $ N '$ set. But here such a problem arises that with a fixed size of $ N $, the set of involved RGF can grow a lot: for example, we glued something in $ N $, the size of $ N $ decreased 10 times, this allowed us to add to $ N $ many new features, which means that for them it is necessary to store the corresponding RGF, thus, the set of stored RGF has increased. In the failed (as it just turned out) model, when there is set $ N '$, the problem was not so pronounced, because at the moment when $ N $ shrinks, we expected that $ N' $ would shrink in the same way due to gluing in $ N '$, which means that some memory for new RGF will be freed.

      From this situation, in my opinion, it is necessary to get out by intentionally throwing out the least relevant RGF from the set of considered (central and peripheral) RGF. However, it is quite possible that here we can get out and still create a structure that stores the involved RGFs, similar to $ N '$, within which it is appropriate to glue something, due to which the amount of memory occupied by the structure will decrease; it is also possible that this structure will allow a combination of this gluing analogue for this structure and discarding the least relevant RGF from consideration.

      An example can be given in which we have a class of features, each of which calculates some function of an element of a finite group obtained from a group unit by sequentially adding the generators of the group and their inverses, in some order (this order is hardwired into TM corresponding to the feature) and possibly some other parameter. If we have added many more such features to $ N $ than the elements in the group contain, then maybe it's time to make a large-scale gluing, dividing such features into classes according to which element of our finite group is the sum of generators and their inverse corresponding to this feature and gluing all the features within each class. And for sure, it will not be necessary for each of these elements to store information about all the features that are glued into this element - it is already clear what element it is (most likely it will be possible to leave information about only a few); so that the amount of used memory is successfully reduced.

      Most likely, it is worth supporting some kind of structure on the set $ N $. For example, it is worth storing information about some propositional constraints on elements of $ N $ that must be satisfied (for example, propositional constraints on the sets of features corresponding to neighboring gates of the original circuit, which say how the gate should be calculated from the gates just below it). It may also be worth storing, for some sets of elements of $ N $ of constant size, the result of applying some propositional formulas to this set. The result of the operation of applying a propositional formula to a constant set of features is obviously also a feature (ideally, such an operation should commute with projection from the version of the set $ N $ before gluing to the version of the set $ N $ after gluing; if it does not commute, this means that you can glue something else).

      Gluing gives me different associations. In the beginning, our consciousness is in a not yet glued form (the $ N $ set coincides with the area in the feature space). Walking through the space of signs, passing through the area of consciousness $ B $, we can notice that we have already seen something like this somewhere (now I am not describing the work of the algorithm): when we walked through the area of consciousness $ A $, such that between its features and features of the area $ B $ exists a natural correspondence, the probability distribution on the features of this area and set of propositional relations betwen them was approximately the same as in the case of the area $ B $. Then we notice the same distribution on the features of the $ C $ area (and same set of propositional relations). And we start to wonder: what if $ A $, $ B $ and $ C $ are actually the same area? It is quite probable that in such a situation it is possible to perform gluing so that the domains $ A $, $ B $ and $ C $ are projected into the same domain of the resulting set $ N $.

Moreover, if the set of propositional connections on sets $ A $, $ B $, $ C $, etc. of equal size (possibly from different floors) is the same, then you can put the same "label" on all the corresponding elements in the sets $ A $, $ B $, $ C $, etc., by doing this for each type of elements (we obtain set of labels equal to the set $ A $). The presence of the same label on two particular features signals to us that these particular features may be a specification of the same, perhaps not considered until now, general feature, though, possibly, with different parameters. The information that certain labels hang on certain features can be used to force the emerging propositional connections (we can force not only equality, but also any other propositional connection - and, by the way, not only propositional) between, firstly, the corresponding features on which these labels are hung, and secondly, between some of the corresponding features on which these labels are hung and some other features that are in the same connections with these features (which are ) - for example, obtained from the first in the same way, and, thirdly, between the features that are in the same connections with some of the corresponding features, on which specific labels are hung (in the third case, the connections between the features which are labeled themselves, or between them and other features, we do not force).

      I have associations between gluing and topological covering: if the set $ N_1 $ is glued into the set $ N_2 $, then $ N_2 $ plays the role of the base of the covering, and $ N_1 $ plays the role of the covering space: several regions from $ N_1 $ can be projected into the еру same region in $ N_2 $ at once. In the case of a (glued) feature space, as in the case of a covering, we can walk in different places and not notice that, in some sense, we are walking in the same place.

      Surely, during the work of the algorithm, there will be situations when, due to some inconsistencies, we understand that we have glued something in vain. You need to learn how to unstick the glued elements in such situations. This is also on the list of things worth working with.

      Important examples of gluing are gluing with zero and gluing with one. When the probabilities at the observed specifications of one general feature correlate with each other, in the sense that a significant majority of them are close to zero, or a significant majority of them are close to one (not to something in between), we glue this feature, respectively, with an identical zero or an identical one, so after that this feature is a constant (and if we use the gluing pressure, then this means that if all the specifications of one RGF are correlated, we start to force their coincidence by imposing the gluing pressure arguments and then gluing the feature with zero or one, and we force the more, the stronger these specifications are correlated - if they hardly correlate, these arguments should be made very weak).

      It is also possible (and quite appropriate) to upgrade our system, in which particular features are divided into two types: those that give out binary values - binary features - and those that give out numerical values (a natural number in the unary number system) - numerical features. Perhaps it would be more convenient to make features that yield several values at once - several binary and several numeric. Numerical values are needed, for example, in order to give them to the input of another feature that calculates something based on the results given by this feature, for example, a pair of numbers can indicate to another feature a specific place in the original circuit.

      It is convenient to consider the place of the tape cell in which the TM head stopped as a natural number produced by the Turing machine.

      The coincidence of two numbers produced by different numerical features gives serious reasons to think about gluing these features in some cases (there is more information in the coincidence of two natural numbers than in the coincidence of two bits).

\par \bigskip
\par \bigskip
\par \bigskip

Further work seems to me, first of all, in the work with gluing: how to efficiently perform gluing, whether is it necessary to maintain some structure on the set $ N $, beyond sets of one or more elements of the feature space projected into each element and a set of propositional clauses - connections between elements ...

      Another important direction is how to build a set of peripheral RGF based on the set of central RGF. I wrote that it might be worth doing just by going through all the small Turing machines and "applying them to the central feature sets" as described above. But it may be worth doing it differently. If a set of central RGF (or a set $ N '$) is given, then what other features would be RELEVANT to consider for possible further addition of their specifications to the set $ N $? This is the question. There are features that are, say so, more or less always relevant. For a string of zeros and ones, it is often relevant to consider a sign that finds the number of ones in such a string. If we consider the entire original circuit - a rectangle, in each cell of which there is a set of zeros and ones, one bit of each type, then it would be relevant to consider a feature that takes two natural numbers - the coordinates of the cell of this circuit and returns the coordinates of the first cell that we meet when moving from a given (given by coordinates) cell straight upwards, in which the bit of the first type is equal to one. The question is what features will be relevant in each specific situation - when we consider a specific set of features, with specific gluings produced. The first thought, as I said, is that the features will be relevant if they are easily (briefly) computable by the set of features already in the field of consideration.

In addition to the small complexity (short computability), there are two points simalteniously that are worth paying attention to when deciding whether to consider some feature relevant. First, it's worth paying attention to the objects that I call "areas of consistency" (perhaps it would be more correct to call such an object a "structure"). The area of consistency is a set of features, for which there are many sufficiently determined propositional relations connecting them with each other, as well as with the already considered features (and added to $ N $). By "sufficiently determined" is meant that the probability of fulfilling of given propositional constraint is close to one. I call them areas of consistency, because they are characterized by the fact that in them it is often possible to predict a specific feature based on different groups of other features, and these predictions are consistent with each other. A feature entering the area of consistensy is often worth considering and adding to consciousness. When we look at a feature and decide whether to add it or not, it is important for us that something else can be said about it, besides the fact that it is calculated according to the features already considered as prescribed by its definition. It is important for us that there is some structure at all.

Secondly, it is worth paying attention to those features for which, when combined with the features already considered, at the current moment of time there is the most discriminated cocyclic polynomial (the more discriminated - the better). This often happens, in particular, when for the same feature there are two strongly contradicting groups of arguments: one prescribes this feature to be one, other - to zero. We consider such signs/groups of signs and "eliminate this tension" - we change the probabilities according to the arguments so as to reduce the discrimination of the most discriminated polynomials (that is, we add these signs to consciousness and work with them in a general manner). We are interested in the groups of features with the most discriminated cocyclic polynomials, since, again, working with them will change the other probabilities for the features under consideration the most. In this aspect, one can recall the work of an investigator who walks through neighbors and asks if they have seen anything strange. We are curious about the strangenesses: when some predictions about an object contradict the predictions about the same object given on other basis. The only difference is that the investigator is looking for the cause of the strangeness, and we are trying to eliminate this strangeness.

Perhaps new features can be "shaped". That is, to represent the scheme that implements the Turing machine in the form of zeros and ones, as described in one of the previous sections, to associate each such bit with the probability of becoming one and modify them together with the rest of the probabilities, among other things, with the help of cocyclic polynomials ... Thus, when shaping / searching for features, we have two goals: firstly, we pursue the structure (areas of consistensy), and secondly, we are looking for the greatest contradiction (tension), in order to eventually eliminate it.

      I suppose that the rich possibility for gluing in the feature space is precisely what distinguishes circuits of small Kolmogorov complexity in a set of general circuits, and perhaps this is something that can help to solve the problem for circuits of small Kolmogorov complexity efficiently in a wide variety of cases.

\subsection {One more idea.}

I'll tell you about one more possible upgrade of the algorithm, for which I, however, am not sure that this will help. First, perhaps, we want to give ourself more freedom and allow the probability distribution on the lamp features to be dependent. I propose for this to consider many - a list - of equal in rights versions of the independent probability distributions on light bulbs and each of them can be changed autonomously (if we talk about one single distribution, then it can be defined as follows: we choose one of the given equitable distributions randomly, equiprobably, and then we choose the arrangement zeros and ones into features (situation) randomly from this distribution; we will call this distribution joint). Moreover, in addition to the cocyclic polynomials discriminated by each specific distribution in the list (and intended to modify this distribution), we will look for cocyclic polynomials discriminated by as many of our distributions as possible and use such polynomials to change all distributions of the list at once. (But it may not work anymore, if we want to find the optimum, as in the case of one independent distribution, simply by finding the minimum of some quadratic function, we need to carry out some kind of joint optimization.)

      The logic of such a construction is that, for example, there may be two lamp features that cannot both be one or both be zero in a correct, consistent (coming from some arrangement of zeros and ones in the basic gates) situation. If we consider one distribution, then at a randomly selected moment in time, it is more likely that either the probabilities for both features will be far from zero and one, or one of the probabilities will be close to one, and the other - to zero. This means that, say, in the case when the first probability is close to zero, and the second - to one, we are looking for cocyclic polynomials "tuned precisely to this special class of situations", that is, discriminated precisely by distributions with such prevailing situations for which the first feature is zero, and the second is one. In the case when we consider many distributions, we will have cocyclic polynomials that "cover" both classes of situations: those in which the first feature is zero, and the second is one, and those in which the first feature is one, and the second is zero.
			
			Since for smaller sets of independent distributions it is expectedly easier to search for cocyclic polynomials discriminated by them jointly, it may make sense to consider a bipartite graph (perhaps some expander), the left part of which is the set of independent distributions, and for each vertex of the right part $ v_i $ we will maintain its own family of cocyclic polynomials $ \Omega_i $, discriminated by the distributions of the left part corresponding to the vertices adjacent to $ v_i $. In this case, each distribution will not be affected by all cocyclic polynomials, but only by those that correspond to the vertices of the right part adjacent to the vertex of the left part corresponding to this distribution.

      Another possible improvement is that, in addition to the current list of feature distributions (let's call it $ A $), we store some of the states of these distributions in past times. At all times that are multiple of a fixed time interval, we remember all the current distributions of the list and add them to another, larger list. We will not change the distributions of this larger list (let's call it $ B $). But we will look for and use, when changing the current probabilities, cocyclic polynomials discriminated by the largest possible number of distributions in the list $ B $.

      Why this is needed can be seen in the example when the list $ A $ consists of only one distribution. If we do not "take screenshots" of this distribution from time to time and send them to the list $ B $, something like a loop can happen: the distribution can fall into some zone in the space of distributions, in which, however, there is no distribution consisting of probabilities zero and one corresponding to the desired (consistent) situation; at the same time, the polynomials we find each time will only shift this distribution within this zone. The list $ B $ is needed just in order to find the polynomials discriminated by all distributions from this zone together, in order to push the current distribution out of this zone.

\subsection {A digression in mathematical logic.}

Here I want to talk about one hypothesis, for which I have no grounds indicating that it should be true, however, if it will turn out that it is true, it would be very interesting, so I decided to add it here.

I am not an expert in mathematical logic, so I ask you to excuse me if this issue in some form has already been investigated, or is easily resolved with the help of some facts already established in this science.

There are a variety of philosophical positions, not all of which imply that every clearly formulated mathematical statement has a specific truth value - yes or no. All this debates about the rule of the excluded middle and the assumption that things like the Continuum Hypothesis may not have a definite yes or no answer at all.

If we try to formulate this hypothesis briefly, then our world may turn out to be only locally consistent, but globally nonconsistent. By local consistency, I mean that any finite set of statements of a given formal system can be associated with a set of truth values of these statements so that they do not contradict each other (I will explain what it means "do not contradict each other" using the example of the formal system described below) ... Global consistency implies that it is possible in a consistent way to associate the truth values to all statements of a given system simaltaneously.

I will work with a formal system close to the feature space described above. It is a set of statements of the form $ \forall x_1 \exists x_2 \forall x_3 ... \exists x_k M (x_1, ..., x_k, y_1, ..., y_l) = 1 $, where $ x_i $ are strings to which quantifiers apply, $ y_i $ - specific strings, own for each statement, $ M $ - a Turing machine that produces a binary value 0 or 1, or does not stop at all, $ k $, $ l $ - specific numbers, own for each statement.

It is possible to protect a certain set of propositional relationships for some finite sets of statements of the form $ \forall x_1 \exists x_2 \forall x_3 ... \exists x_k M (x_1, ..., x_k, y_1, ..., y_l) = 1 $. To protect, or to make a propositional connection protected, is to prove that a given set of statements cannot take a given set of truth values. On this system of statements, one can absolutely in the same way as it was in the space of features to consider light bulbs, some of which are protected (those that, when lit, imply a protected propositional connection), and some are not. With any choice of binary truth values for statements, some bulbs light up and some do not, each according to its own internal propositional law, everything is exactly the same as in the case of the feature space, with the only difference that now there are infinitely many statements (and light bulbs).

In order to prove the global inconsistency, it is sufficient to provide a protected cocyclic absolutely convergent series, the existence of which is exactly what my hypothesis predicts. This series is a function from a set of bulbs into a set of real numbers, the set of values of which, firstly, is an absolutely converging series, and secondly, the sum of the values of the bulbs that are lit does not depend on the choice of binary values for each of the statements of the system and is equal to one, and thirdly, positive values can be associated only with protected bulbs. It can be thought about whether it is possible to make the number of statements that form a light bulb from the set of light bulbs we are considering, not limited from above. This time, the invariance of such a series will be proven, perhaps in a different way from how it was in the case of cocyclic polynomials. Perhaps by counting the sum of certain series.
 
It is clear that if a protected cocyclic absolutely convergent series exists, then this proves that it is impossible to simultaneously assign truth values 0 and 1 to all statements of the indicated type so that they do not contradict each other (do not form protected propositional connections).


\begin{thebibliography}{99}

\bibitem{GP2007} 
B. Goertzel, C. Pennachin (Eds.):
Artificial General Intelligence.
XVI, 509 pages. 2007

\bibitem{LP}
Л. Г. Хачиян, “Полиномиальный алгоритм в линейном программировании”, Докл. АН СССР, 244:5 (1979),  1093–1096  

\bibitem{Shap}
Шаповалов А. В.
Ш24 Принцип узких мест. — 2-е изд., доп. — М.: МЦНМО,
2008. — 32 с.: ил.

\end{thebibliography}
\end{document}